\documentclass[a4paper]{article}

\usepackage{INTERSPEECH2021}

\usepackage{cite}
\usepackage{amsmath,amssymb,amsfonts}
\usepackage{algorithmic}
\usepackage{graphicx}
\usepackage{textcomp}
\usepackage{multirow}
\usepackage{hrefhide}
\usepackage{mathtools}

\newcommand*{\dittoclosing}{ \raisebox{-0.5ex}{''} }

\usepackage{color, colortbl}
\definecolor{LightCyan}{rgb}{0.88,1,1}

\usepackage[first=0,last=9]{lcg}

\begin{document}
\title{COVID-19 Detection in Cough, Breath and Speech using Deep Transfer Learning and Bottleneck Features}

\name{Madhurananda Pahar$^1$, Marisa Klopper$^2$, Robin Warren$^2$ and Thomas Niesler$^1$}
\address{
	$^1$Department of Electrical and Electronic Engineering, Stellenbosch University, South Africa \\
	$^2$SAMRC Centre for Tuberculosis Research, Division of Molecular Biology and Human Genetics, Faculty of Medicine and Health Sciences, Stellenbosch University, South Africa}
\email{$^1$\{mpahar, trn\}@sun.ac.za, $^2$\{marisat, rw1\}@sun.ac.za}

\maketitle

\begin{abstract}
	We present an experimental investigation into the effectiveness of transfer learning and bottleneck feature extraction in detecting COVID-19 from audio recordings of cough, breath and speech.
	This type of screening is non-contact, does not require specialist medical expertise or laboratory facilities and can be deployed on inexpensive consumer hardware.
	We use datasets that contain recordings of coughing, sneezing, speech and other noises, but do not contain COVID-19 labels, to pre-train three deep neural networks: a CNN, an LSTM and a Resnet50.
	These pre-trained networks are subsequently either fine-tuned using smaller datasets of coughing with COVID-19 labels in the process of transfer learning, or are used as bottleneck feature extractors.
	Results show that a Resnet50 classifier trained by this transfer learning process delivers optimal or near-optimal performance across all datasets achieving areas under the receiver operating characteristic (ROC AUC) of 0.98, 0.94 and 0.92 respectively for all three sound classes (coughs, breaths and speech).
	This indicates that coughs carry the strongest COVID-19 signature, followed by breath and speech.
	Our results also show that applying transfer learning and extracting bottleneck features using the larger datasets without COVID-19 labels led not only to improve performance, but also to minimise the standard deviation of the classifier AUCs among the outer folds of the leave-$p$-out cross-validation, indicating better generalisation.
	We conclude that deep transfer learning and bottleneck feature extraction can improve COVID-19 cough, breath and speech audio classification, yielding automatic classifiers with higher accuracy. 
\end{abstract}



\section{Introduction}\label{sec:introduction}

COVID-19 (\textbf{CO}rona \textbf{VI}rus \textbf{D}isease of 20\textbf{19}) was declared a global pandemic on February 11, 2020 by the World Health Organisation (WHO). 
Caused by the severe acute respiratory syndrome coronavirus 2 (SARS-CoV-2), this disease affects the respiratory system and includes symptoms like fatigue, dry cough, shortness of breath, joint pain, muscle pain, gastrointestinal symptoms and loss of smell or taste \cite{carfi2020persistent,wang2020clinical}. 
Due to its effect on the vascular endothelium, the acute respiratory distress syndrome can originate from either the gas or vascular side of the alveolus which becomes visible in a chest x-ray or CT scan for COVID-19 patients~\cite{marini2020management, aguiar2020inside}. 
Among the patients infected with SARS-CoV-2, between 5\% and 20\% are admitted to ICU and their mortality rate varies between 26\% and 62\% \cite{ziehr2020respiratory}. 
Medical lab tests are available to diagnose COVID-19 by analysis of exhaled breaths \cite{davis2021breath}. 
This technique was reported to achieve an accuracy of 93\% when considering a group of 28 COVID-19 positive and 12 COVID-19 negative patients~\cite{grassin2021metabolomics}. 
Related work using a group of 25 COVID-19 positive and 65 negative patients achieved an area under the ROC curve (AUC) of 0.87~\cite{ruszkiewicz2020diagnosis}. 

Previously, machine learning algorithms have been applied to detect COVID-19 using image analysis. 
For example, COVID-19 was detected from computed tomography (CT) images using a Resnet50 architecture with 96.23\% accuracy in \cite{walvekar2020detection}.
The same architecture also detected pneumonia due to COVID-19 with an accuracy of 96.7\%~\cite{sotoudeh2020artificial} and COVID-19 from x-ray images with an accuracy of 96.30\% \cite{yildirim2020deep}. 

The automatic analysis of cough audio for COVID-19 detection has also received attention.
Coughing is a predominant symptom of many lung ailments and its effect on the respiratory system
varies~\cite{higenbottam2002chronic,  chang2008chronic}. 
Lung disease can cause the glottis to behave differently and the airway to be either restricted or obstructed and this can influence the acoustics of the vocal audio such as cough, breath and speech~\cite{chung2008prevalence, knocikova2008wavelet}.
This raises the prospect of identifying the coughing audio associated with a particular respiratory disease such as COVID-19~\cite{imran2020ai4covid, laguarta2020covid}. 
Researchers have found that a simple binary machine learning classifier can distinguish between healthy and COVID-19 respiratory audio, such as coughs gathered from crowdsourced data, with AUC higher than 0.8 \cite{brown2020exploring}.
Improved performance was achieved using a convolutional neural network (CNN) for cough and breath audio, achieving an AUC of 0.846~\cite{coppock2021end}. 

In our previous work, we have also found that automatic COVID-19 detection is possible on the basis of the acoustic cough signal~\cite{pahar2020covid}. 
Here we extend this work firstly by considering whether breath and speech audio can also be used effectively for COVID-19 detection.
Secondly, since the COVID-19 datasets at our disposal are comparatively small, we apply transfer learning and extract bottleneck features to take advantage of other datasets that do not include COVID-19 labels. 
To do this, we use both publicly available and our own datasets to pre-train three deep neural network (DNN) architectures: a CNN, a long short-term memory (LSTM) and a 50-layer residual based architecture (Resnet50). 
For subsequent COVID-19 classifier evaluation, the Coswara dataset~\cite{sharma2020coswara}, the Interspeech Computational Paralinguistics ChallengE (ComParE) dataset~\cite{Schuller21-TI2} and the Sarcos dataset \cite{pahar2020covid} are used.
We report further evidence of accurate discrimination using all three audio classes and conclude that vocal audio including coughing, breathing and speech are all affected by the condition of the lungs to an extent that they carry acoustic information that can be used by machine learning classifiers to detect signatures of COVID-19.
We find that the application of transfer learning enables the classifiers to perform more accurately, exhibit a greater robustness and reduce the tendency of over-fitting.

Sections~\ref{sec:data} and Section~\ref{sec:features} summarise the datasets used for experimentation and the primary feature extraction process. 
Section \ref{sec:transferlearning} describes the transfer learning process and Section~\ref{sec:bottleneck} explains the bottleneck feature extraction process. 
Section \ref{sec:evaluation} presents the experimental setup, including the cross-validated hyperparameter optimisation and classifier evaluation process. 
Experimental results are presented in Section~\ref{sec:results} and discussed in Section \ref{sec:discussion}.
Finally, Section~\ref{sec:conclusion} summarises and concludes this study.

\section{Data}
\label{sec:data}

\subsection{Datasets without COVID-19 labels for pre-training}
\label{sec:pretraindata}

\begin{table*}[ht!]
	\footnotesize
	\renewcommand\arraystretch{1.4}
	\setlength\minrowclearance{1.0pt}
	\caption{\textbf{Summary of the Datasets used in Pre-training.} Classifiers are pre-trained on 10.29 hours audio recordings annotated with four class labels: cough, sneeze, speech and noise. The datasets do not include any COVID-19 labels. }  
	\centering 
	\begin{center}
		\begin{tabular}{ c | c | c | c | c | c | c }
			\hline
			\hline
			\textbf{Type} & \textbf{Dataset} & \textbf{Sampling Rate} & \textbf{No of Events} & \textbf{Total audio} & \textbf{Average length} & \textbf{Standard deviation}  \\
			\hline
			\hline
			\multirow{5}{*}{Cough} & TASK dataset & 44.1 kHz & 6000 & 91 mins & 0.91 sec & 0.25 sec  \\
			\cline{2-7}
			& Brooklyn dataset & 44.1 kHz & 746 & 6.29 mins & 0.51 sec & 0.21 sec  \\
			\cline{2-7}
			& Wallacedene dataset  & 44.1 kHz & 1358 & 17.42 mins & 0.77 sec & 0.31 sec  \\
			\cline{2-7}
			& Google Audio Set \& Freesound & 16 kHz & 3098 & 32.01 mins & 0.62 sec & 0.23 sec \\
			\cline{2-7}
			& \textbf{Total (Cough)} & \textbf{---} & \textbf{11202} & \textbf{2.45 hours} & \textbf{0.79 sec} & \textbf{0.23 sec} \\
			
			\hline \hline
			\multirow{3}{*}{Sneeze} & Google Audio Set \& Freesound & 16 kHz & 1013 & 13.34 mins & 0.79 sec & 0.21 sec  \\
			\cline{2-7}
			& Google Audio Set \& Freesound + SMOTE & 16 kHz & 9750 & 2.14 hours & 0.79 sec & 0.23 sec \\
			\cline{2-7}
			& \textbf{Total (Sneeze)} & \textbf{---} & \textbf{10763} & \textbf{2.14 hours} & \textbf{0.79 sec} & \textbf{0.23 sec} \\
			
			\hline \hline
			\multirow{3}{*}{Speech} & Google Audio Set \& Freesound & 16 kHz & 2326 & 22.48 mins & 0.58 sec & 0.14 sec  \\
			\cline{2-7}
			& LibriSpeech & 16 kHz & 56 & 2.54 hours & 2.72 mins & 0.91 mins  \\
			\cline{2-7}
			& \textbf{Total (Speech)} & \textbf{---} & \textbf{2382} & \textbf{2.91 hours} & \textbf{4.39 sec} & \textbf{0.42 sec} \\
			
			\hline \hline
			\multirow{3}{*}{Noise}& TASK dataset & 44.1 kHz & 12714 & 2.79 hours & 0.79 sec & 0.23 sec \\
			\cline{2-7}
			& Google Audio Set \& Freesound & 16 kHz & 1027 & 11.13 mins & 0.65 sec & 0.26 sec \\
			\cline{2-7}
			& \textbf{Total (Noise)} & \textbf{---} & \textbf{13741} & \textbf{2.79 hours} & \textbf{0.79 sec} & \textbf{0.23 sec} \\
			\hline
			\hline
			
		\end{tabular}
	\end{center}
	\label{table:pre-train-dataset-summary}
\end{table*}

Audio data with COVID-19 labels remain scarce which limits classifier training.
We have therefore made use of five additional datasets without COVID-19 labels for pre-training.
These datasets contain recordings of coughing, sneezing, speech and non-vocal audio. 
The first three datasets (TASK, Brooklyn and Wallacedene) were compiled by ourselves as part of research projects concerning cough monitoring and cough classification.
The last two (Google Audio Set \& Freesound and Librispeech) were compiled from publicly available data.
Since all five datasets were compiled before the start of the COVID-19 pandemic, they are unlikely to contain data from COVID-19 positive subjects.
All datasets used for pre-training include manual annotations. 

\subsubsection{TASK dataset}
This corpus consists of spontaneous coughing audio collected at a small tuberculosis (TB) clinic near Cape Town, South Africa~\cite{pahar2021deep}.
The dataset contains 6000 cough audio by patients undergoing TB treatment and 11393 non-cough audio such as laughter, doors opening and objects moving. 
This data was intended for the development of cough detection algorithms and the recordings were made in a multi-ward environment using a smartphone with an external microphone.

\subsubsection{Brooklyn dataset}
This dataset contains 746 voluntary coughs by 38 subjects compiled for the development of cough audio classification systems~\cite{botha2018detection}.
Audio recording took place in a controlled indoor booth, using a Rode M3 microphone and an audio field recorder.

\subsubsection{Wallacedene dataset}
This dataset consists of 1358 voluntary coughs by 51 patients, also compiled for the development of cough audio classification~\cite{pahar2021tb}.
In this case, audio recording took place in an outdoor booth located at a busy primary healthcare clinic.
The recording was performed using a Rode M1 microphone and an audio field recorder. 
This data has more environmental noise and therefore a poorer signal-to-noise ratio than the Brooklyn dataset.

\subsubsection{Google Audio Set \& Freesound}
The Google Audio Set dataset contains excerpts from 1.8 million Youtube videos that have been manually labelled according to an ontology of 632 audio event categories~\cite{gemmeke2017audio}. 
The Freesound audio database is a collection of tagged sounds uploaded by contributors from around the world ~\cite{font2013freesound}. 
In both datasets, the audio recordings were contributed by many different individuals under widely varying recording conditions and noise levels. 
From these two datasets, we have compiled a collection of 3098 coughing audio, 1013 sneezing audio, 2326 speech excerpts and 1027 other non-vocal audio such as engine noise, running water and restaurant chatter.
Previously, this dataset was used for the development of cough detection algorithms~\cite{miranda2019comparative}. 

\subsubsection{LibriSpeech}
As a source of speech audio data, we have selected utterances by 28 male and 28 female speakers from the freely available LibriSpeech corpus~\cite{panayotov2015librispeech}. 
These recordings contain very little noise and the large size of the corpus allowed easy gender balancing.

\subsubsection{Summary of data used for pre-training}

In total, the data described above includes 11202 cough events (2.45 hours of audio). 
It also includes 2.91 hours of speech from both male and female participants and 2.98 hours of other non-vocal audio.
Finally, the data also includes recordings of 1013 sneezing audio, totalling 13.34 minutes of audio.
Hence sneezing is under-represented as a class in the pre-training data.
Since such an imbalance in training data can detrimentally affect the performance especially of neural networks~\cite{van2007experimental,krawczyk2016learning}, we have applied the synthetic minority over-sampling technique (SMOTE)~\cite{chawla2002smote}. 
SMOTE oversamples the minor class by creating additional synthetic samples rather than, for example, random oversampling. 
We have in the past successfully applied SMOTE to address training set class imbalances in cough detection~\cite{pahar2021deep} and cough classification~\cite{pahar2020covid} based on audio recordings. 

In total, therefore, a dataset containing  10.29 hours of audio recordings annotated with four class labels was available to pre-train the neural architectures.
The composition of this dataset is summarised in Table~\ref{table:pre-train-dataset-summary}. 
All recordings used for pre-training were downsampled at 16 kHz. 

\subsection{Datasets with COVID-19 labels for classification}
\label{sec:coviddata}

Three datasets of coughing audio with COVID-19 labels were available for experimentation.

\begin{table*}[ht!]
	\footnotesize
	\renewcommand\arraystretch{1.4}
	\setlength\minrowclearance{1.0pt}
	\caption{\textbf{Summary of the datasets used for COVID-19 classification.} Cough, breath and speech signals were extracted from the Coswara, ComParE and Sarcos datasets. COVID-19 positive subjects are under-represented in all three datasets.}  
	\centering 
	\begin{center}
		\begin{tabular}{ c | c | c | c | c | c | c | c }
			\hline
			\hline
			\textbf{Type} & \textbf{Dataset} & \textbf{Sampling Rate} & \textbf{Label} & \textbf{Subjects} & \textbf{Total audio} & \textbf{Average per subject } & \textbf{Standard deviation }  \\
			
			\hline
			\hline
			
			\multirow{9}{*}{Cough} & \multirow{3}{*}{Coswara} & \multirow{3}{*}{44.1 kHz} & COVID-19 Positive & 92 & 4.24 mins & 2.77 sec & 1.62 sec \\
			\cline{4-8}
			&  &  & Healthy & 1079 & 0.98 hours & 3.26 sec & 1.66 sec  \\
			\cline{4-8}
			&  &  & \textbf{Total} & \textbf{1171} & \textbf{1.05 hours} & \textbf{3.22 sec} & \textbf{1.67 sec}  \\
			\cline{2-8}
			
			& \multirow{3}{*}{ComParE} & \multirow{3}{*}{16 kHz} & COVID-19 Positive & 119 & 13.43 mins & 6.77 sec & 2.11 sec \\
			\cline{4-8}
			&  &  & Healthy & 398 & 40.89 mins & 6.16 sec & 2.26 sec  \\
			\cline{4-8}
			&  &  & \textbf{Total} & \textbf{517} & \textbf{54.32 mins} & \textbf{6.31 sec} & \textbf{2.24 sec}  \\
			\cline{2-8}
			
			& \multirow{3}{*}{Sarcos} & \multirow{3}{*}{44.1 kHz} & COVID-19 Positive & 18  & 0.87 mins & 2.91 sec & 2.23 sec \\
			\cline{4-8}
			&  &  & COVID-19 Negative & 26 & 1.57 mins & 3.63 sec & 2.75 sec \\
			\cline{4-8}
			&  &  & \textbf{Total} & \textbf{44} & \textbf{2.45 mins} & \textbf{3.34 sec} & \textbf{2.53 sec} \\
			
			\hline
			\hline
			
			\multirow{3}{*}{Breath} & \multirow{3}{*}{Coswara} & \multirow{3}{*}{44.1 kHz} & COVID-19 Positive & 88 & 8.58 mins & 5.85 sec & 5.05 sec \\
			\cline{4-8}
			&  &  & Healthy & 1062 & 2.77 hours & 9.39 sec & 5.23 sec  \\
			\cline{4-8}
			&  &  & \textbf{Total} & \textbf{1150} & \textbf{2.92 hours} & \textbf{9.126 sec} & \textbf{5.29 sec}  \\
			
			\hline
			\hline
			
			\multirow{9}{*}{Speech} & \multirow{2.5}{*}{Coswara} & \multirow{3}{*}{44.1 kHz} & COVID-19 Positive & 88 & 12.42 mins & 8.47 sec & 4.27 sec \\
			\cline{4-8}
			& \multirow{2}{*}{(normal count)} &  & Healthy & 1077 & 2.99 hours & 9.99 sec & 3.09 sec  \\
			\cline{4-8}
			&  &  & \textbf{Total} & \textbf{1165} & \textbf{3.19 hours} & \textbf{9.88 sec} & \textbf{3.22 sec}  \\
			\cline{2-8}
			
			& \multirow{2.5}{*}{Coswara} & \multirow{3}{*}{44.1 kHz} & COVID-19 Positive & 85 & 7.62 mins & 5.38 sec & 2.76 sec \\
			\cline{4-8}
			& \multirow{2}{*}{(fast count)} &  & Healthy & 1074 & 1.91 hours & 6.39 sec & 1.77 sec  \\
			\cline{4-8}
			&  &  & \textbf{Total} & \textbf{1159} & \textbf{2.03 hours} & \textbf{6.31 sec} & \textbf{1.88 sec}  \\
			\cline{2-8}
			
			& \multirow{3}{*}{ComParE} & \multirow{3}{*}{16 kHz} & COVID-19 Positive & 214 & 44.02 mins & 12.34 sec & 5.35 sec \\
			\cline{4-8}
			&  &  & Healthy & 396 & 1.46 hours & 13.25 sec & 4.67 sec  \\
			\cline{4-8}
			&  &  & \textbf{Total} & \textbf{610} & \textbf{2.19 hours} & \textbf{12.93 sec} & \textbf{4.93 sec}  \\
			
			\hline
			\hline
			
		\end{tabular}
	\end{center}
	\label{table:class-dataset-summary}
\end{table*}

\subsubsection{Coswara dataset}

This dataset 
is specifically developed with the testing of classification algorithms for COVID-19 detection in mind. 
Data collection is web-based, and participants contribute by using their smartphones to record their coughing, breathing and speech. 
Audio 
recordings were collected of both shallow and deep breaths as well as speech uttered at a normal and fast pace. 
However, since the deep breaths consistently outperformed the shallow breaths in our initial experiments, the latter will not be presented in our experiments. 
At the time of writing, the data included contributions from participants located on five different continents~\cite{sharma2020coswara, pahar2020covid,muguli2021dicova}. 

\begin{figure}[h!]
	\centerline{\includegraphics[width=0.51\textwidth]{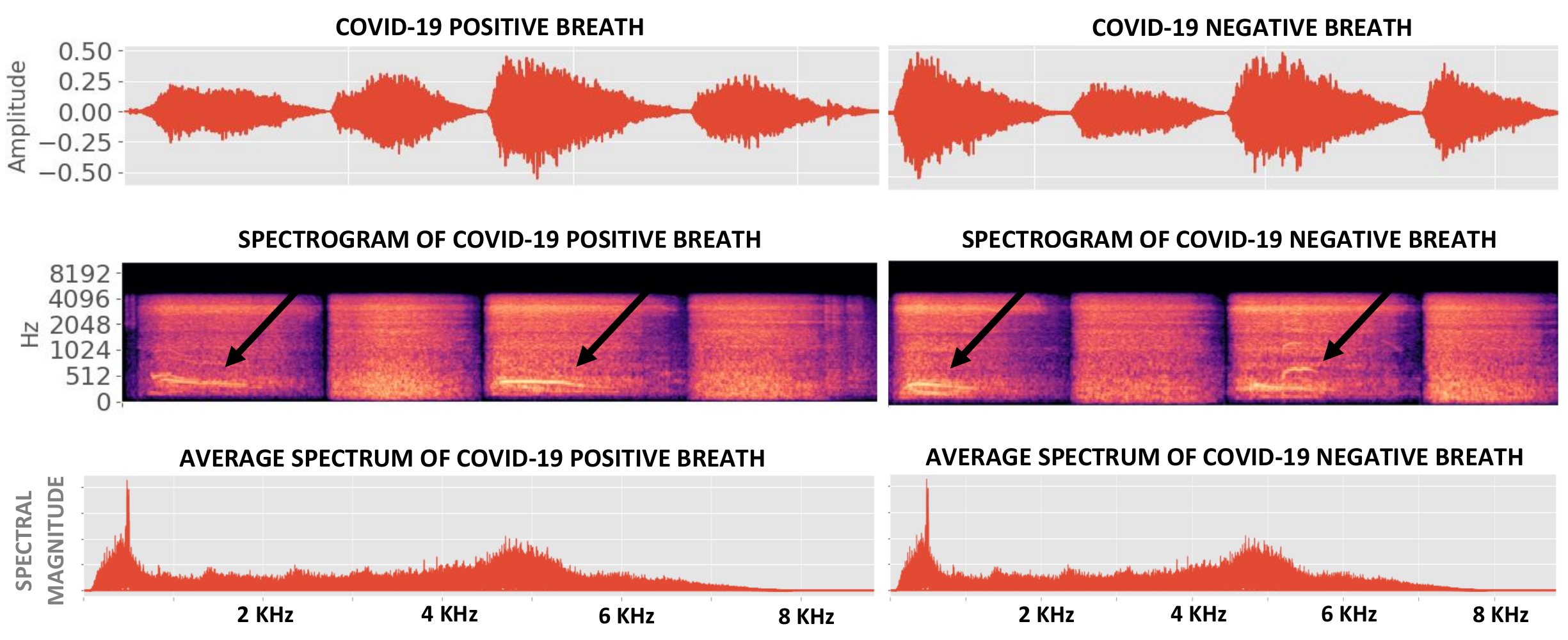}}
	\caption{\textbf{Pre-processed breath} signals from both COVID-19 positive and COVID-19 negative subjects in the Coswara dataset. Breaths corresponding to inhalation are marked by arrows, and are followed by an exhalation. 
	}
	\label{fig:processed-breath-info}
\end{figure}

\begin{figure}[h!]
	\centerline{\includegraphics[width=0.51\textwidth]{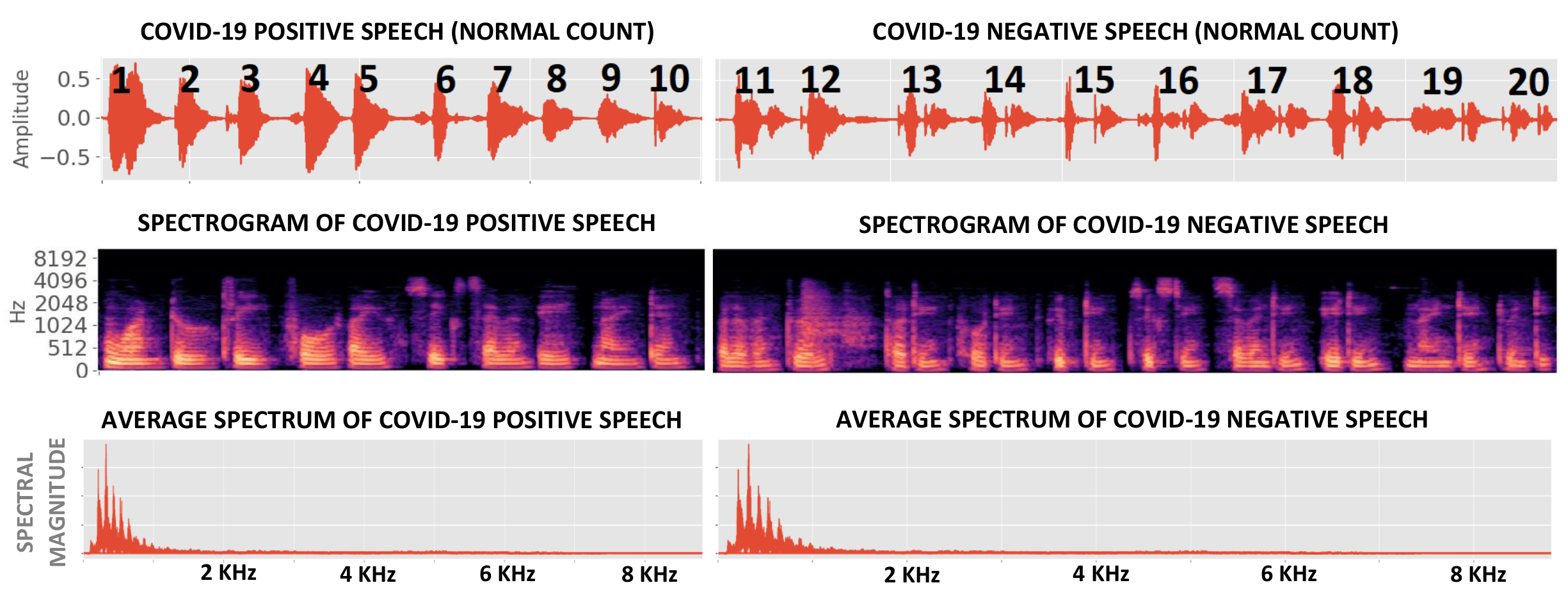}}
	\caption{\textbf{Pre-processed speech} (counting from 1 to 20 at a normal pace) from both COVID-19 positive and COVID-19 negative subjects in the Coswara dataset. In contrast to breath (Figure~\ref{fig:processed-breath-info}), spectral energy is concentrated below 1 kHz.
	}
	\label{fig:processed-nor-count-info}
\end{figure}

Figure~\ref{fig:processed-breath-info} and Figure~\ref{fig:processed-nor-count-info} show examples of Coswara breaths and normal-paced count from one to twenty respectively, collected from both COVID-19 positive and COVID-19 negative subjects. 
It is evident that breaths have more higher-frequency content than speech and interesting to note that COVID-19 breaths are 30\% shorter than non-COVID-19 breaths (Table~\ref{table:class-dataset-summary}). 
All audio recordings were pre-processed to remove periods of silence to within a margin of 50 ms using a simple energy detector.

\subsubsection{ComParE dataset}

This dataset was provided as a part of the 2021 Interspeech Computational Paralinguistics ChallengE (ComParE)~\cite{Schuller21-TI2}. 
The ComParE dataset contains recordings, 
of both coughs and speech, where the latter is the utterance `I hope my data can help to manage the virus pandemic' in the speaker's language of choice. 

\subsubsection{Sarcos dataset}
This dataset was collected in South Africa as part of this research and currently contains recordings of coughing by 18 COVID-19 positive and 26 COVID-19 negative subjects. 
Audio was 
pre-processed in the same way as the Coswara data.
Since this dataset is very small, we have used it in our previous work exclusively for an independent validation~\cite{pahar2020covid}.
In this study, however, it has also been used to fine-tune and evaluate pre-trained DNN classifiers by means of transfer learning and the extraction of bottleneck features. 

\subsubsection{Summary of data used for classification}
A summary of the above three datasets is presented in Table~\ref{table:class-dataset-summary}.
We see that the COVID-19 positive class is under-represented in all cases.
To address this, we again apply SMOTE during training. 
We also note that the Coswara dataset contains the largest number of subjects, followed by ComParE and then Sarcos and 
all recordings were downsampled at 16 kHz.

\section{Primary Feature Extraction }
\label{sec:features}

From the time-domain audio signals, we have extracted mel-frequency cepstral coefficients (MFCCs) and linearly-spaced log filterbank energies, along with their respective velocity and acceleration coefficients.
We have also extracted the signal zero-crossing rate (ZCR) and kurtosis~\cite{bachu2010voiced}, which are indicative respectively of time-domain signal variability and tailedness i.e. the prevalence of higher amplitudes. 

MFCCs have been very effective mostly in speech \cite{pahar_coding_2020}, 
but also in 
discriminating dry and wet coughs \cite{chatrzarrin2011feature}, 
and recently in characterising COVID-19 audio~\cite{alsabek2020studying}. 
Linearly-spaced log filterbank energies have proved useful in biomedical applications, including cough audio classification~\cite{aydin2009log, botha2018detection, pahar2021tb}. 
The ZCR is the number of times the time-domain signal changes sign within a frame, and is an indicator of variability~\cite{bachu2010voiced}.

\begin{figure}[h!]
	\centerline{\includegraphics[width=0.51\textwidth]{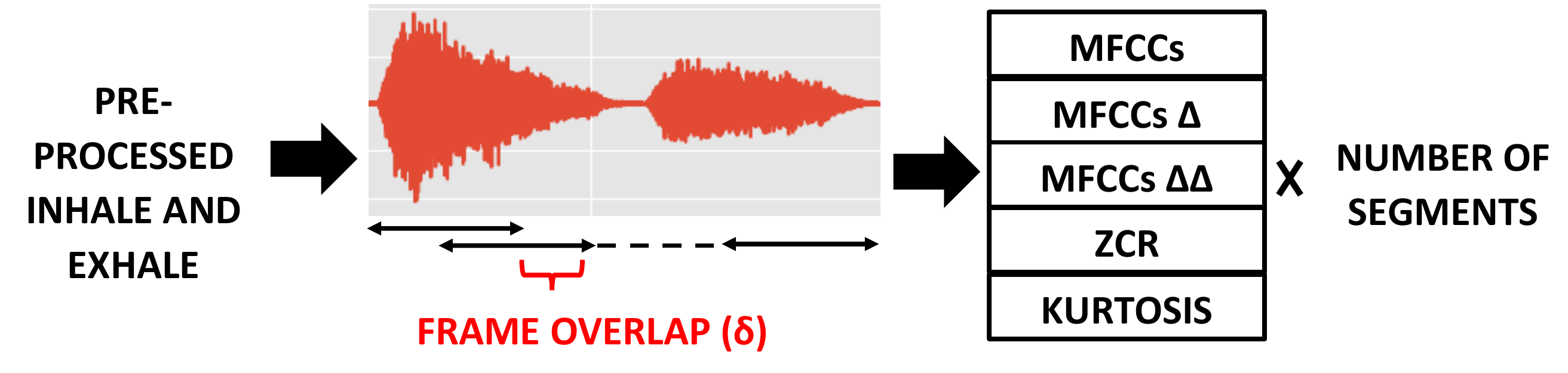}}
	\caption{\textbf{Feature extraction process for a breath audio.} The frame overlap $\delta$ is adjusted in such a way that the entire recording is divided into $\mathcal{S}$ segments. For $\mathcal{M}$ MFCCs, the final feature matrix has ($3\mathcal{M} + 2, \mathcal{S}$) dimensions. }
	\label{fig:feat-extract}
\end{figure}

Features are extracted from overlapping frames, where the frame overlap $\delta$ is computed to ensure that the audio signal is always divided into exactly $\mathcal{S}$ frames, as illustrated in Figure~\ref{fig:feat-extract}.
This ensures that the entire audio event is always captured within a fixed number of frames, which allows a fixed input dimension to be maintained while preserving the general overall temporal structure of the sound.
Such fixed two-dimensional feature dimensions are particularly useful for the training of DNN classifiers, and has been found to perform well in previous experiments \cite{pahar2020covid}.

The frame length ($\mathcal{F}$), number of frames ($\mathcal{S}$), number of lower order MFCCs ($\mathcal{M}$) and number of linearly spaced filters ($\mathcal{B}$) are regarded as feature extraction hyperparameters, listed in Table~\ref{table:feat-hyper-parameter}. 
The table shows that in our experiments each audio signal is divided into between 70 and 200 frames, each between 512 and 4096 samples i.e. 32 msec and 256 msec long.
The number of extracted MFCCs ($\mathcal{M}$) lies between 13 and 65, and the number of linearly-spaced filterbanks ($\mathcal{B}$)  between 40 and 200.
This allows the spectral information included in each feature to be varied.


\begin{table}[h!]
	\footnotesize
	\renewcommand\arraystretch{1.4}
	\setlength\minrowclearance{1.0pt}
	\caption{\textbf{Primary feature (PF) extraction hyperparameters.} We have used between 13 and 65 MFCCs and between 40 and 200 linearly spaced filters to extract log energies. } 
	\centering 
	\begin{center}
		\begin{tabular}{ c | c | c }
			\hline
			\hline
			\textbf{Hyperparameters} & \textbf{Description}     & \textbf{Range} \\
			
			\hline
			\multirow{2}{*}{MFCCs ($\mathcal{M}$)}   & lower order    & $13 \times k$, where \\[-0.2em]
			& MFCCs to keep            & $k=1, 2, 3, 4, 5$ \\ 
			
			\hline
			Linearly spaced         & used to extract   & 40 to 200 \\[-0.2em]
			filters ($\mathcal{B}$) &  log energies & in steps of 20 \\
			
			\hline
			\multirow{2}{*}{Frame length ($\mathcal{F}$)}  & into which audio  & $2^k$ where \\[-0.2em]
			&  is segmented    &  $k=9, 10, 11, 12$\\
			
			\hline
			\multirow{2}{*}{Segments ($\mathcal{S}$)}    & number of frames & $10 \times k$, where \\[-0.2em]
			&  extracted from audio  & $k=7, 10, 12, 15, 20$ \\
			
			\hline
			\hline
		\end{tabular}
	\end{center}
	\label{table:feat-hyper-parameter}
\end{table}

The input feature matrix to the classifiers has the dimension of ($3\mathcal{M} + 2, \mathcal{S}$) for $\mathcal{M}$ MFCCs along with their $\mathcal{M}$ velocity and $\mathcal{M}$ acceleration coefficients, as shown in Figure \ref{fig:feat-extract}. 
Similarly, for linearly spaced filters, the dimension of the feature matrix is ($3\mathcal{B} + 2, \mathcal{S}$). 


We will refer to the features described in this section as \textbf{primary features} (PF) to distinguish them from the bottleneck features (BNF), described in Section~\ref{sec:bottleneck}.


\section{Transfer Learning architecture}
\label{sec:transferlearning}

Since the audio datasets with COVID-19 labels described in Section~\ref{sec:coviddata} are small, they may lead to overfitting when training deep architectures.
Nevertheless, in previous work we have found that deep architectures perform better than shallow classifiers when using these as training sets \cite{pahar2020covid}. 
In this work, we consider whether the classification performance of such DNNs can be improved by applying transfer learning.

To achieve this, we use the datasets containing 10.29 hours of audio, labelled with four classes: cough, sneeze, speech and noise, but do not include COVID-19 labels (Table~\ref{table:pre-train-dataset-summary} in Section~\ref{sec:pretraindata}).
This data is used to pre-train three deep neural architectures: a CNN, an LSTM and a Resnet50. 
The feature extraction hyperparameters: $\mathcal{M}=39, \mathcal{F}=2^{10}$ and $\mathcal{S}=150$ delivered good performance in our previous work \cite{pahar2020covid} and thus have also been used here (Table \ref{table:pre-train-hyper-parameter}). 

The CNN consists of three convolutional layers, with 256, 128 and 64 (2$\times$2) kernels respectively and each followed by (2,2) max-pooling.
The LSTM consists of three layers with 512, 256 and 128 LSTM units respectively, each including dropout with a rate of 0.2.
A standard Resnet50, as described in Table~1 of~\cite{he2016deep}, has been implemented with 512-dimensional dense layers. 

During pre-training, all three networks (CNN, LSTM and Resnet50) are terminated by three dense layers with dimensionalities 512, 64 and finally 4 to correspond to the four classes mentioned in Table \ref{table:pre-train-dataset-summary}. 
Relu activation functions were used throughout, except in the four-dimensional output layer which was softmax.
All the above architectural hyperparameters were chosen by optimising the four-class classifiers during cross-validation (Table \ref{table:pre-train-hyper-parameter}).

\begin{table}[h]
	\footnotesize
	\renewcommand\arraystretch{1.4}
	\setlength\minrowclearance{1.0pt}
	\caption{\textbf{Hyperparameters of the pre-trained networks:} Feature extraction hyperparameters were adopted from the optimal values in previous related work~\cite{pahar2020covid}, while classifier hyperparameters were optimised on the pre-training data using cross-validation. }
	\centering 
	\begin{center}
		\begin{tabular}{ c | c | c  }

			\hline
			\hline
			\multicolumn{3}{c}{\textbf{\uppercase{Feature Extraction hyperparameters}}}\\
			\hline
			\hline
			
			\multicolumn{2}{c|}{\textbf{Hyperparameters}} & \textbf{Values} \\
			
			\hline
			$\mathcal{M}$   & MFCCs & $39$ \\
			
			\hline
			$\mathcal{F}$  & Frame length  & $2^{10} = 1024$ \\
			
			\hline
			$\mathcal{S}$  & Segments & $150$ \\

			\hline
			\hline
			\multicolumn{3}{c}{\textbf{\uppercase{Classifier hyperparameters}}}\\
			\hline
			\hline
			\textbf{Hyperparameters} & \textbf{Classifier} & \textbf{Values} \\
			\hline
			Conv filters & CNN & $256$ \& $128$ \& $64$ \\
			\hline
			Kernel size & CNN & $2$ \\
			\hline
			Dropout rate & CNN, LSTM  & $0.2$ \\
			\hline
			Dense layer & \multirow{2}{*}{CNN, LSTM, Resnet50} & \multirow{2}{*}{$512$ \& $64$ \& $4$} \\[-0.2em]
			(while pre-training) &  &  \\
			\hline
			Dense layer & \multirow{2}{*}{CNN, LSTM, Resnet50} & \multirow{2}{*}{$32$ \& $2$} \\[-0.2em]
			(while fine-tuning) &  &  \\
			\hline
			LSTM units & LSTM & $512$ \& $256$ \& $128$ \\
			\hline
			Learning rate & LSTM & $10^{-3} = 0.001$ \\
			\hline
			Batch Size & CNN, LSTM, Resnet50  & $2^7 = 128$ \\
			\hline
			Epochs & CNN, LSTM, Resnet50 & $70$ \\
			\hline
			\hline
		\end{tabular}
	\end{center}
	\label{table:pre-train-hyper-parameter}
\end{table}

\begin{figure*}[h!]
	\centerline{\includegraphics[width=0.99\textwidth]{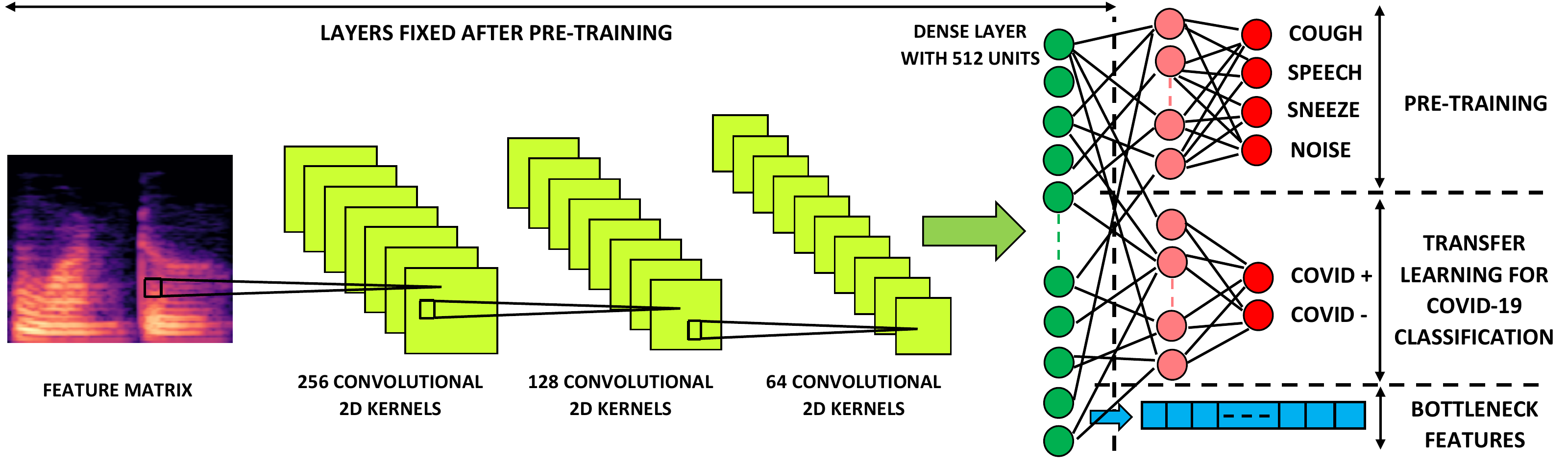}}
	\caption{\textbf{Transfer Learning Architecture for a CNN.} Cross-validation on the pre-training data determined the optimal CNN architecture to have three convolutional layers with 256, 128 and 64 (2$\times$2) kernels respectively and each followed by (2,2) max-pooling. The convolutional layers were followed by two dense layers with 512 and 64 relu units each, and the network was terminated by a 4-dimensional softmax.  To apply transfer learning, the final two layers were removed and replaced with a new dense layer and a terminating 2-dimensional softmax to account for COVID-19 positive and negative classes. Only this newly added portion of the network was trained for classification on the data with COVID-19 labels. In addition, the outputs of the third-last layer (512-dimensional dense relu) from the pre-trained network were used as bottleneck features. }
	\label{fig:class-arch}
\end{figure*}

After pre-training on the datasets described in Section~\ref{sec:pretraindata}, the 64 and 4-dimensional dense layers terminating the network were discarded from the CNN, the LSTM and the Resnet50.
This left three trained deep neural networks, each accepting the same input dimensions and each with a 512-dimensional relu output layer.
The parameters of these three pre-trained networks were then fixed for the remaining experiments.

In order to obtain COVID-19 classifiers by transfer learning, two dense layers are added after the 512-dimensional output layer of each of the three pre-trained deep networks.
The final layer is a two-dimensional softmax, to indicate COVID-19 positive and negative classes respectively.
The dimensionality of the penultimate layer was also considered to be a hyperparameter and was optimised during nested k-fold cross-validation.
Its optimal value was found to be 32 for all three architectures.
The transfer learning process for a CNN architecture is illustrated in Figure~\ref{fig:class-arch}.


\section{Bottleneck Features}
\label{sec:bottleneck}

The 512-dimensional output of the three pre-trained networks described in the previous section has a much lower dimensionality than the ($3\mathcal{M}+2$, $\mathcal{S}$) i.e. $(3 \times 39 + 2)\times150 = 17850$ dimensional input matrix consisting of primary features (Table \ref{table:pre-train-hyper-parameter}).
Therefore, the output of this layer can be viewed as a bottleneck feature vector \cite{silnova2018but, song2015deep, nguyen2013optimizing}. 
In addition to fine-tuning, where we add terminating dense layers to the three pre-trained networks and optimise these for the binary COVID-19 detection task as shown in Figure~\ref{fig:class-arch}, we have trained logistic regression (LR), support vector machine (SVM), k-nearest neighbour (KNN) and multilayer perceptron (MLP) classifiers using these bottleneck features as inputs.
Bottleneck features computed by the CNN, the LSTM or the Resnet50 were chosen based on the one which performed better in the corresponding transfer learning experiments.
Since the Resnet50 achieved higher development set AUCs than the CNN and the LSTM during transfer learning, it was used to extract bottleneck features on which the LR, SVM, KNN and MLP classifiers were trained.

\section{Experimental Method}\label{sec:evaluation}

We have evaluated the effectiveness of transfer learning (Section~\ref{sec:transferlearning}) and bottleneck feature extraction (Section~\ref{sec:bottleneck}) using CNN, LSTM and Resnet50 architectures in improving the performance of COVID-19 classification based on cough, breath and speech audio signals.
In order to place these results in context, we provide two baselines.

\begin{enumerate}
	\item As a first baseline, we train the three deep architectures (CNN, LSTM and Resnet50) directly on the primary features extracted from data containing COVID-19 labels (as described in Section~\ref{sec:coviddata}) and hence skip the pre-training. Some of these baseline results were developed in our previous work \cite{pahar2020covid}. 
	\item As a second baseline, we train shallow classifiers (LR, SVM, KNN and MLP) on the primary input features (as described in Section~\ref{sec:features}), also extracted from the data containing COVID-19 labels (described in Section~\ref{sec:coviddata}). 
\end{enumerate}

The performance of these baseline systems will be compared against:

\begin{enumerate}
	\item Deep architectures (CNN, LSTM and Resnet50) trained by the transfer learning process. The respective deep architectures are pre-trained (as described in Section~\ref{sec:transferlearning}), after which the final two layers are fine-tuned on the data containing COVID-19 labels (as described in Section~\ref{sec:coviddata}). 
	\item Shallow architectures (LR, SVM, KNN and MLP) trained on the bottleneck features extracted from the pre-trained networks.
\end{enumerate}

\subsection{Hyperparameter Optimisation}

Hyperparameters for three pre-trained networks have already been described in Section~\ref{sec:transferlearning} and are listed in Table \ref{table:pre-train-hyper-parameter}. 
The remaining hyperparameters are those of the baseline deep classifiers (CNN, LSTM and Resnet50 without pre-training), the four shallow classifiers (LR, SVM, KNN and MLP), and the dimensionality of the penultimate layer for the deep architectures during transfer learning. 



With the exception of Resnet50, all these hyperparameters optimisation and performance evaluation has been performed within the inner loops of a leave-$p$-out nested cross-validation scheme~\cite{liu2019leave}.
Due to the excessive computational requirements of optimising Resnet50 metaparameters within the same cross-validation framework, we have used the standard 50 skip layers in all experiments \cite{he2016deep}. 
Classifier hyperparameters and the values considered during optimisation are listed in Table~\ref{table:class-hyper-parameter}.
A five-fold split, similar to that employed in~\cite{pahar2020covid}, was used for the nested leave-$p$-out cross-validation.

\begin{table}[h]
	\footnotesize
	\renewcommand\arraystretch{1.4}
	\setlength\minrowclearance{1.0pt}
	\caption{\textbf{Classifier hyperparameters}, optimised using leave-$p$-out nested cross-validation. }
	\centering 
	\begin{center}
		\begin{tabular}{ l | c | l  }
			\hline
			\hline
			\textbf{Hyperparameters} & \textbf{Classifier} & \textbf{Range} \\
			\hline
			\hline
			Regularisation & \multirow{2}{*}{LR, SVM} & $10^i$ where, \\[-0.2em]
			Strength ($\alpha_1$) &  & $i=-7,-6,\ldots,6,7$ \\
			\hline
			$l1$ penalty ($\alpha_2$)  & LR  & 0 to 1 in steps of 0.05 \\
			\hline
			$l2$ penalty ($\alpha_3$)  & LR, MLP  & 0 to 1 in steps of 0.05 \\
			\hline
			Kernel & \multirow{2}{*}{SVM} & $10^i$ where, \\[-0.2em]
			Coefficient ($\alpha_4$) &  & $i=-7,-6,\ldots,6,7$ \\ 
			\hline
			No. of neighbours ($\alpha_5$) & KNN & 10 to 100 in steps of 10 \\
			\hline
			Leaf size ($\alpha_6$) & KNN & 5 to 30 in steps of 5 \\
			\hline
			No. of neurons ($\alpha_7$) & MLP & 10 to 100 in steps of 10 \\
			\hline
			No. of conv filters ($\beta_1$) & CNN & $3 \times 2^k$ where $k=3, 4, 5$ \\
			\hline
			Kernel size ($\beta_2$) & CNN & 2 and 3 \\
			\hline
			Dropout rate ($\beta_3$) & CNN, LSTM  & 0.1 to 0.5 in steps of 0.2 \\
			\hline
			Dense layer size ($\beta_4$) & CNN, LSTM  & $2^k$ where $k=4, 5$ \\
			\hline
			LSTM units ($\beta_5$) & LSTM & $2^k$ where $k=6, 7, 8$ \\
			\hline
			\multirow{2}{*}{Learning rate ($\beta_6$)} & \multirow{2}{*}{LSTM, MLP} & $10^k$ where, \\[-0.2em]
			&  &  $k=-2,-3,-4$ \\ 
			\hline
			Batch Size ($\beta_7$) & CNN, LSTM  & $2^k$ where $k=6, 7, 8$\\
			\hline
			Epochs ($\beta_8$) & CNN, LSTM  & 10 to 250 in steps of 20 \\
			\hline
			\hline
		\end{tabular}
	\end{center}
	\label{table:class-hyper-parameter}
\end{table}

\subsection{Classifier Evaluation}
Receiver operating characteristic (ROC) curves were calculated within both the inner and outer loops of the leave-$p$-out cross-validation scheme described in the previous section. 
The inner-loop ROC values were used for the hyperparameter optimisation, while the average of the outer-loop ROC values indicate final classifier performance on the independent held-out test sets.
The AUC score indicates how well the classifier has performed over a range of decision thresholds \cite{fawcett2006introduction}. 
The threshold that achieves an equal error rate ($\gamma_{EE}$) was computed from these curves. 

We note the mean per-frame probability that an event such as a cough is from a COVID-19 positive subject by $\hat{P}$:
\begin{equation}
	\hat{P} = \frac{\sum\limits_{i=1}^{K} P(Y = 1|X_i, \theta)}{K} 
	\label{eq:P-hat}
\end{equation} 
where $K$ indicates the number of frames in the cough and $P(Y = 1|X_i, \theta)$ is the output of the classifier for feature vector $X_i$ and parameters $\theta$ for the $i^{th}$ frame. 
Now we define the indicator variable $C$ as:
\begin{equation}
	C = 
	\begin{cases}
		1 & \text{if } \hat{P}\geq \gamma_{EE}\\
		0              & \text{otherwise}
	\end{cases}
	\label{eq:C-sum}
\end{equation}

We then define two COVID-19 index scores as $\text{CI}_1$ and $\text{CI}_2$, 
$N_1$ as the number of coughs from the subject in the recording and $N_2$ as the total number of frames of cough audio gathered from the subject in Equations~\ref{eq:COVIDI-1} and~\ref{eq:COVIDI-2} respectively. 

\begin{equation}
	\text{CI}_1 = \frac{\sum\limits_{i=1}^{N_1} C}{N_1} 
	\label{eq:COVIDI-1}
\end{equation}

\begin{equation}
	\text{CI}_2 = \frac{\sum\limits_{i=1}^{N_2} P(Y = 1|X_i)}{N_2} 
	\label{eq:COVIDI-2}
\end{equation}

Hence Equation \ref{eq:P-hat} computes a per-cough average probability while Equation~\ref{eq:COVIDI-2} computes a per-frame average probability. 
For the shallow classifiers, the use of one of Equations~\ref{eq:COVIDI-1} and~\ref{eq:COVIDI-2} was considered an additional hyperparameter during cross-validation, and it was found that taking the maximum value of the index scores consistently led to the best performance. 

The average specificity, sensitivity and accuracy, as well as the AUC together with its standard deviation ($\sigma_{AUC}$) are shown in Tables~\ref{table:cough-classifier-results}, \ref{table:breath-classifier-results} and~\ref{table:speech-classifier-results} for cough, breath and speech respectively.
These values have all been calculated over the outer folds during nested cross-validation. 
Hyperparameters producing the highest AUC over the inner loops have been noted as the `best classifier hyperparameter'.

\section{Experimental Results}
\label{sec:results}

COVID-19 classification performance based on cough, breath and speech are presented in Tables~\ref{table:cough-classifier-results}, \ref{table:breath-classifier-results} and \ref{table:speech-classifier-results} respectively. 
These tables include the performance of baseline deep classifiers without pre-training, deep classifiers trained by transfer learning (TL), shallow classifiers using bottleneck features (BNF) and baseline shallow classifiers trained directly on the primary features (PF). 
The best performing classifiers appear first for each dataset and the baseline results are shown towards the end.
Each system is identified by an `ID'. 

\subsection{Coughs}

\begin{table*}[!h]
	\scriptsize
	\renewcommand\arraystretch{1.4}
	\setlength\minrowclearance{1.0pt}
	\setlength{\tabcolsep}{5pt} 
	\centering
	\caption{\textbf{COVID-19 cough classifier performance:} For the Coswara, Sarcos and ComParE datasets the highest AUCs were 0.982, 0.961 and 0.944 respectively and achieved by a Resnet50 trained by transfer learning in the first two cases and a KNN classifier using 12 primary features determined by sequential forward selection (SFS) in the third case. When Sarcos as a validation-only datset for a classifier trained on the Coswara data, an AUC of 0.954 is achieved.
	}
	\begin{tabular}{ c | c | c | c | c | c | c | c | c | c }
		\hline
		\hline
		{\multirow{2}{*}{\textbf{Dataset}}} & 
		{\multirow{2}{*}{\textbf{ID}}} & 
		{\multirow{2}{*}{\textbf{Classifier}}} & \textbf{Best Feature} &
		\textbf{Best Classifier Hyperparameters} & \multicolumn{5}{c}{\textbf{Performance}} \\
		
		\cline{6-10}
		
		&  &  & \textbf{Hyperparameters} & \textbf{(Optimised inside nested cross-validation)} & \textbf{Spec} & \textbf{Sens} & \textbf{Acc} & \textbf{AUC} & \textbf{$\sigma_{AUC}$} \\
		
		\hline
		\hline
		\multirow{10}{*}{Coswara} & \textit{C1} & \textit{Resnet50+TL} & \textit{Table \ref{table:pre-train-hyper-parameter}} & \textit{Default Resnet50 (Table 1 in \cite{he2016deep})} & \textit{97\%} & \textit{98\%} & \textit{97\%} & \textit{\textbf{0.982}} & \textit{2$\times 10^{-3}$} \\
		\cline{2-10}
		& C2 & CNN+TL & \dittoclosing & Table \ref{table:pre-train-hyper-parameter} & 92\% & 98\% & 95\% & 0.972 & 3$\times 10^{-3}$ \\
		\cline{2-10}
		& C3 & LSTM+TL & \dittoclosing & \dittoclosing & 93\% & 95\% & 94\% & 0.964 & 3$\times 10^{-3}$ \\
		\cline{2-10}
		& C4 & MLP+BNF & \dittoclosing & $\alpha_3$=0.35, $\alpha_7$=50 & 92\% & 96\% & 94\% & 0.963 & 4$\times 10^{-3}$ \\
		\cline{2-10}
		& C5 & SVM+BNF & \dittoclosing & $\alpha_1=10^{4}$, $\alpha_4=10^{1}$ & 89\% & 93\% & 91\% & 0.942 & 3$\times 10^{-3}$ \\
		\cline{2-10}
		& C6 & KNN+BNF & \dittoclosing & $\alpha_5$=20, $\alpha_6$=15 & 88\% & 90\% & 89\% & 0.917 & 7$\times 10^{-3}$ \\
		\cline{2-10}
		& C7 & LR+BNF & \dittoclosing & $\alpha_1=10^{-1}$, $\alpha_2=0.5$, $\alpha_3=0.5$ & 84\% & 86\% & 85\% & 0.898 & 8$\times 10^{-3}$ \\
		\cline{2-10}
		& C8 & Resnet50+PF \cite{pahar2020covid} & Table 4 in \cite{pahar2020covid} & Default Resnet50 (Table 1 in \cite{he2016deep}) & 98\% & 93\% & 95\% & 0.976 & 18$\times 10^{-3}$ \\
		\cline{2-10}
		& C9 & CNN+PF \cite{pahar2020covid} & \dittoclosing & Table 4 in \cite{pahar2020covid} & 99\% & 90\% & 95\% & 0.953 & 39$\times 10^{-3}$ \\
		\cline{2-10}
		& C10 & LSTM+PF \cite{pahar2020covid} & \dittoclosing & \dittoclosing & 97\% & 91\% & 94\% & 0.942 & 43$\times 10^{-3}$ \\
		
		\hline
		
		\multirow{7}{*}{Sarcos} & \textit{C11} & \textit{Resnet50+TL} & Table \ref{table:pre-train-hyper-parameter} & \textit{Default Resnet50 (Table 1 in \cite{he2016deep})} & \textit{92\%} & \textit{96\%} & \textit{94\%} & \textit{0.961} & \textit{3$\times 10^{-3}$} \\
		\cline{2-10}
		& C12 & LSTM+TL & \dittoclosing & Table \ref{table:pre-train-hyper-parameter} & 92\% & 92\% & 92\% & 0.943 & 3$\times 10^{-3}$ \\
		\cline{2-10}
		& C13 & CNN+TL & \dittoclosing & \dittoclosing & 89\% & 91\% & 90\% & 0.917 & 4$\times 10^{-3}$ \\
		\cline{2-10}
		& C14 & MLP+BNF & \dittoclosing & $\alpha_3$=0.75, $\alpha_7$=70 & 88\% & 90\% & 89\% & 0.913 & 7$\times 10^{-3}$ \\
		\cline{2-10}
		& C15 & SVM+BNF & \dittoclosing & $\alpha_1=10^{-2}$, $\alpha_4=10^{4}$ & 88\% & 89\% & 89\% & 0.904 & 6$\times 10^{-3}$ \\
		\cline{2-10}
		& C16 & KNN+BNF & \dittoclosing & $\alpha_5$=40, $\alpha_6$=20 & 85\% & 87\% & 86\% & 0.883 & 8$\times 10^{-3}$ \\
		\cline{2-10}
		& C17 & LR+BNF & \dittoclosing & $\alpha_1=10^{-3}$, $\alpha_2=0.4$, $\alpha_3=0.6$ & 83\% & 86\% & 85\% & 0.867 & 9$\times 10^{-3}$ \\
		
		\hline
		
		\multirow{2.5}{*} {Sarcos} & \textit{C18} & \textit{Resnet50+TL} & \dittoclosing & \textit{Default Resnet50 (Table 1 in \cite{he2016deep})} & \textit{92\%} & \textit{96\%} & \textit{94\%} & \textit{0.954} & --- \\
		\cline{2-10}
		
		\multirow{2}{*} {(val only)} & C19 & LSTM+PF \cite{pahar2020covid} & Table 5 in \cite{pahar2020covid} & Table 5 in \cite{pahar2020covid} & 73\% & 75\% & 74\% & 0.779 & --- \\
		\cline{2-10}
		
		& C20 & LSTM+PF+SFS \cite{pahar2020covid} & \dittoclosing & \dittoclosing & 96\% & 91\% & 93\% & 0.938 & --- \\
		
		\hline

		\multirow{12}{*}{ComParE} & C21 & Resnet50+TL & Table \ref{table:pre-train-hyper-parameter} & Default Resnet50 (Table 1 in \cite{he2016deep}) & 89\% & 93\% & 91\% & 0.934 & 4$\times 10^{-3}$ \\
		\cline{2-10}
		& C22 & LSTM+TL & \dittoclosing & Table \ref{table:pre-train-hyper-parameter} & 88\% & 92\% & 90\% & 0.916 & 4$\times 10^{-3}$ \\
		\cline{2-10}
		& C23 & CNN+TL & \dittoclosing & \dittoclosing & 86\% & 90\% & 88\% & 0.898 & 4$\times 10^{-3}$ \\
		\cline{2-10}
		& C24 & MLP+BNF & \dittoclosing & $\alpha_3$=0.25, $\alpha_7$=20 & 85\% & 90\% & 88\% & 0.912 & 5$\times 10^{-3}$ \\
		\cline{2-10}
		& C25 & SVM+BNF & \dittoclosing & $\alpha_1=10^{-3}$, $\alpha_4=10^{2}$ & 85\% & 90\% & 88\% & 0.903 & 6$\times 10^{-3}$ \\
		\cline{2-10}
		& C26 & KNN+BNF & \dittoclosing & $\alpha_5$=70, $\alpha_6$=20 & 85\% & 86\% & 86\% & 0.882 & 8$\times 10^{-3}$ \\
		\cline{2-10}
		& C27 & LR+BNF & \dittoclosing & $\alpha_1=10^{4}$, $\alpha_2=0.3$, $\alpha_3=0.7$ & 84\% & 86\% & 85\% & 0.863 & 8$\times 10^{-3}$ \\
		\cline{2-10}
		& \textit{C28} & \textit{KNN+PF+SFS} & \textit{$\mathcal{B}=60, \mathcal{F}=2^{11}, \mathcal{S}=70$} & \textit{$\alpha_5$=60, $\alpha_6$=25} & \textit{84\%} & \textit{90\%} & \textit{92\%} & \textit{0.944} & \textit{9$\times 10^{-3}$} \\
		\cline{2-10}
		& C29 & KNN+PF & $\mathcal{B}=60, \mathcal{F}=2^{11}, \mathcal{S}=70$ & $\alpha_5$=60, $\alpha_6$=25 & 78\% & 80\% & 80\% & 0.855 & 13$\times 10^{-3}$ \\
		\cline{2-10}
		& C30 & MLP+PF & $\mathcal{M}=13, \mathcal{F}=2^{10}, \mathcal{S}=100$ & $\alpha_3$=0.65, $\alpha_7$=40 & 76\% & 80\% & 78\% & 0.839 & 14$\times 10^{-3}$ \\
		\cline{2-10}
		& C31 & SVM+PF & $\mathcal{B}=80, \mathcal{F}=2^{9}, \mathcal{S}=70$ & $\alpha_1=10^{-4}$, $\alpha_4=10^{-1}$ & 75\% & 78\% & 77\% & 0.814 & 12$\times 10^{-3}$ \\
		\cline{2-10}
		& C32 & LR+PF & $\mathcal{B}=140, \mathcal{F}=2^{11}, \mathcal{S}=70$ & $\alpha_1=10^{-2}$, $\alpha_2=0.6$, $\alpha_3=0.4$ & 69\% & 73\% & 71\% & 0.789 & 13$\times 10^{-3}$ \\
		
		\hline
		\hline
		
	\end{tabular}
	\label{table:cough-classifier-results}
\end{table*}

We have found in our previous work \cite{pahar2020covid} that, when training a Resnet50 on only the Coswara dataset, an AUC of 0.976 ($\sigma_{AUC} = 0.018$) can be achieved for the binary classification problem of distinguishing COVID-19 coughs from healthy coughs. 
These results are reproduced as baseline systems C8, C9 and C10 in Table~\ref{table:cough-classifier-results}. 
The improved results achieved by transfer learning are indicated by systems C1 to C7 in the same table.
Specifically, system C1 shows that, by applying transfer learning as described in Section~\ref{sec:transferlearning}, the same Resnet50 architecture can achieve an AUC of 0.982 ($\sigma_{AUC} = 0.002$). 
The entries for systems C2 and C3 show that pre-training also improves the AUCs achieved by the deep CNN and LSTM classifiers from 0.953 (system C9) to 0.972 (system C2) and from 0.942 (system C10) to 0.964 (system C3) respectively. 
Of particular note in all these cases is the substantial decrease in the standard deviation of the AUC ($\sigma_{AUC}$), observed during cross-validation when implementing transfer learning. 
This indicates that pre-training leads to classifiers with more consistent performance on the unseen test data. 


The Sarcos dataset is much smaller than the Coswara dataset and too small to train a deep classifier directly.
For this reason, it was used only as an independent validation dataset for classifiers trained on the Coswara data in our previous work~\cite{pahar2020covid}.
It can however be used to fine-tune pre-trained classifiers during transfer learning, and the resulting performance is reflected by systems C11 to C17 in Table~\ref{table:cough-classifier-results}. 
Previously an AUC of 0.938 (system C20) was achieved when using Sarcos as an independent validation data and applying sequential forward selection (SFS).
Here, we find that transfer learning applied to the Resnet50 model results in an AUC of 0.961 (system C11) and a lower standard deviation ($\sigma_{AUC}=$0.003) than that observed for the Coswara dataset \cite{pahar2020covid}. 
As an additional experiment, we apply the Resnet50 classifier trained by transfer learning using the Coswara data to the Sarcos data, thus again using the latter as an independent validation set.
The resulting performance is indicated by system C18, while the previous baselines are repeated as systems C19 and C20. 
System C18 achieves an AUC of 0.954, which is only slightly below the 0.961, achieved by system C11 where the pre-trained model used the Sarcos data for fine-tuning, and slightly higher than the AUC of 0.938 achieved by system C20 which is the baseline LSTM trained on Coswara without transfer learning but employing SFS~\cite{devijver1982pattern}.
This supports our earlier observation that transfer learning appears to lead to more robust classifiers that can generalise to other datasets.
Due to the extreme computational load, we have not yet been able to evaluate SFS within the transfer learning framework. 

For the ComParE dataset, we have included shallow classifiers trained directly on the primary input features (KNN+PF, MLP+PF, SVM+PF and LR+PF).
These are the baseline systems C29 to C32 in Table~\ref{table:cough-classifier-results}. 
The best-performing shallow classifier is C29, where a KNN used 60 linearly spaced filterbank log energies as features. 
System C28 is the result of applying SFS to system C29. 
In this case, SFS identifies the top 12 features based on the development sets used during nested cross-validation, and results in the best-performing shallow system with an AUC of 0.944. 
This represents a substantial improvement over the AUC of 0.855 achieved by the same system without SFS (system C29). 
Systems C21 to C27 in Table~\ref{table:cough-classifier-results} are obtained by transfer learning using the ComParE dataset. 
These show improved performance over the shallow classifiers without SFS. 
In particular, after transfer learning, the Resnet50 achieves almost the same AUC as the best ComParE system (system C28) with a lower $\sigma_{AUC}$. 

\begin{figure}[h!]
	\centerline{\includegraphics[width=0.5\textwidth]{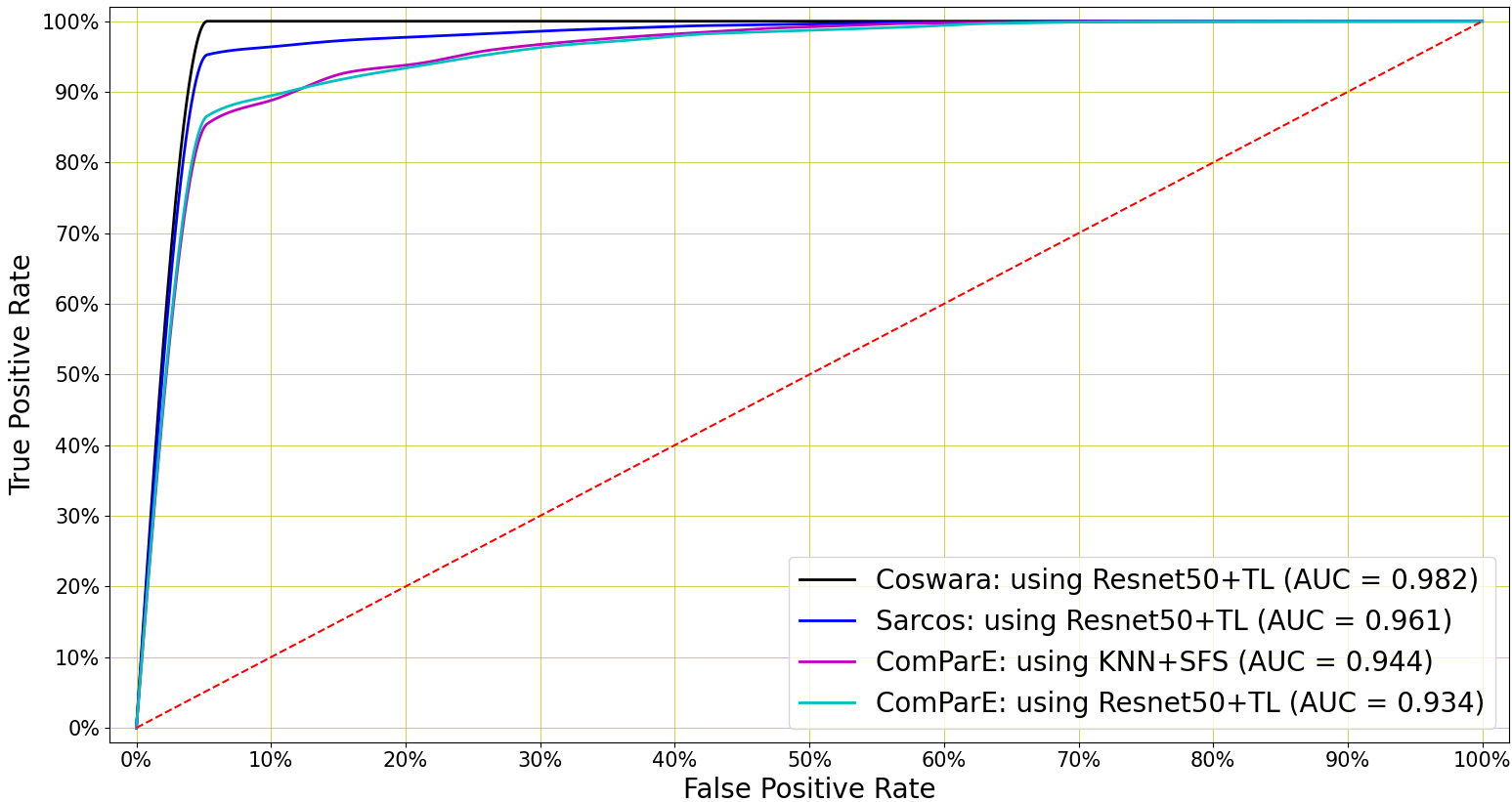}}
	\caption{\textbf{COVID-19 cough classification}: A Resnet50 classifier with transfer learning achieved the highest AUC in classifying COVID-19 coughs for the Coswara and Sarcos datasets (0.982 and 0.961 respectively). For the ComParE dataset,  AUCs of 0.944 and 0.934 were achieved by a KNN classifier using 12 features identified by SFS and by a Resnet50 classifier trained by transfer learning respectively. }
	\label{fig:cough-result}
\end{figure}

When considering the performance of the shallow classifiers trained on the bottleneck features across all three datasets in Table~\ref{table:cough-classifier-results}, we see that a consistent improvement over the use of primary features with the same classifiers is observed. 
The ROC curves for the best-performing COVID-19 cough classifiers are shown in Figure \ref{fig:cough-result}.

\subsection{Breath}

\begin{table*}[!h]
	\scriptsize
	\renewcommand\arraystretch{1.4}
	\setlength\minrowclearance{1.0pt}
	\setlength{\tabcolsep}{5pt} 
	\centering
	\caption{\textbf{COVID-19 breath classifier performance:} For breaths, the best performance was achieved by an SVM using bottleneck features (AUC = 0.942). The Resnet50 classifier trained by transfer learning achieves a similar AUC of 0.934. 
	}
	\begin{tabular}{ c | c | c | c | c | c | c | c | c | c }
		\hline
		\hline
		{\multirow{2}{*}{\textbf{Dataset}}} &
		{\multirow{2}{*}{\textbf{ID}}} &
		{\multirow{2}{*}{\textbf{Classifier}}} & \textbf{Best Feature} &
		\textbf{Best Classifier Hyperparameters} & \multicolumn{5}{c}{\textbf{Performance}} \\
		
		\cline{6-10}
		
		&  &  & \textbf{Hyperparameters} & \textbf{(Optimised inside nested cross-validation)} & \textbf{Spec} & \textbf{Sens} & \textbf{Acc} & \textbf{AUC} & \textbf{$\sigma_{AUC}$} \\
		
		\hline
		\hline
		\multirow{10}{*}{Coswara} & B1 & Resnet50+TL & Table \ref{table:pre-train-hyper-parameter} & Default Resnet50 (Table 1 in \cite{he2016deep}) & 87\% & 93\% & 90\% & 0.934 & 3$\times 10^{-3}$ \\
		\cline{2-10}
		& B2 & LSTM+TL & \dittoclosing & Table \ref{table:pre-train-hyper-parameter} &86\% & 90\% & 88\% & 0.927 & 3$\times 10^{-3}$ \\
		\cline{2-10}
		& B3 & CNN+TL & \dittoclosing & \dittoclosing & 85\% & 89\% & 87\% & 0.914 & 3$\times 10^{-3}$ \\
		\cline{2-10}
		& \textit{B4} & \textit{SVM+BNF} & \dittoclosing & \textit{$\alpha_1=10^{2}$, $\alpha_4=10^{-2}$} & \textit{88\%} & \textit{94\%} & \textit{91\%} & \textit{\textbf{0.942}} & \textit{4$\times 10^{-3}$} \\
		\cline{2-10}
		& B5 & MLP+BNF & \dittoclosing & $\alpha_3$=0.45, $\alpha_7$=50 & 87\% & 93\% & 90\% & 0.923 & 6$\times 10^{-3}$ \\
		\cline{2-10}
		& B6 & KNN+BNF & \dittoclosing & $\alpha_5$=70, $\alpha_6$=10 & 87\% & 93\% & 90\% & 0.922 & 9$\times 10^{-3}$ \\
		\cline{2-10}
		& B7 & LR+BNF & \dittoclosing & $\alpha_1=10^{-4}$, $\alpha_2=0.8$, $\alpha_3=0.2$ & 86\% & 90\% & 88\% & 0.891 & 8$\times 10^{-3}$ \\
		\cline{2-10}
		& B8 & Resnet50+PF & $\mathcal{M}=39, \mathcal{F}=2^{10}, \mathcal{S}=150$ & Default Resnet50 (Table 1 in \cite{he2016deep}) & 92\% & 90\% & 91\% & 0.923 & 34$\times 10^{-3}$ \\
		\cline{2-10}
		& B9 & LSTM+PF & $\mathcal{M}=26, \mathcal{F}=2^{11}, \mathcal{S}=120$ & $\beta_3$=0.1, $\beta_4$=32, $\beta_5$=128, $\beta_6$=0.001, $\beta_7$=256, $\beta_8$=170 & 90\% & 86\% & 88\% & 0.917 & 41$\times 10^{-3}$ \\
		\cline{2-10}
		& B10 & CNN+PF & $\mathcal{M}=52, \mathcal{F}=2^{10}, \mathcal{S}=100$ & $\beta_1$=48, $\beta_2$=2, $\beta_3$=0.3, $\beta_4$=32, $\beta_7$=256, $\beta_8$=210 & 87\% & 85\% & 86\% & 0.898 & 42$\times 10^{-3}$ \\
		\cline{3-10}
		
		\hline
		\hline
		
	\end{tabular}
	\label{table:breath-classifier-results}
\end{table*}

\begin{figure}[h!]
	\centerline{\includegraphics[width=0.5\textwidth]{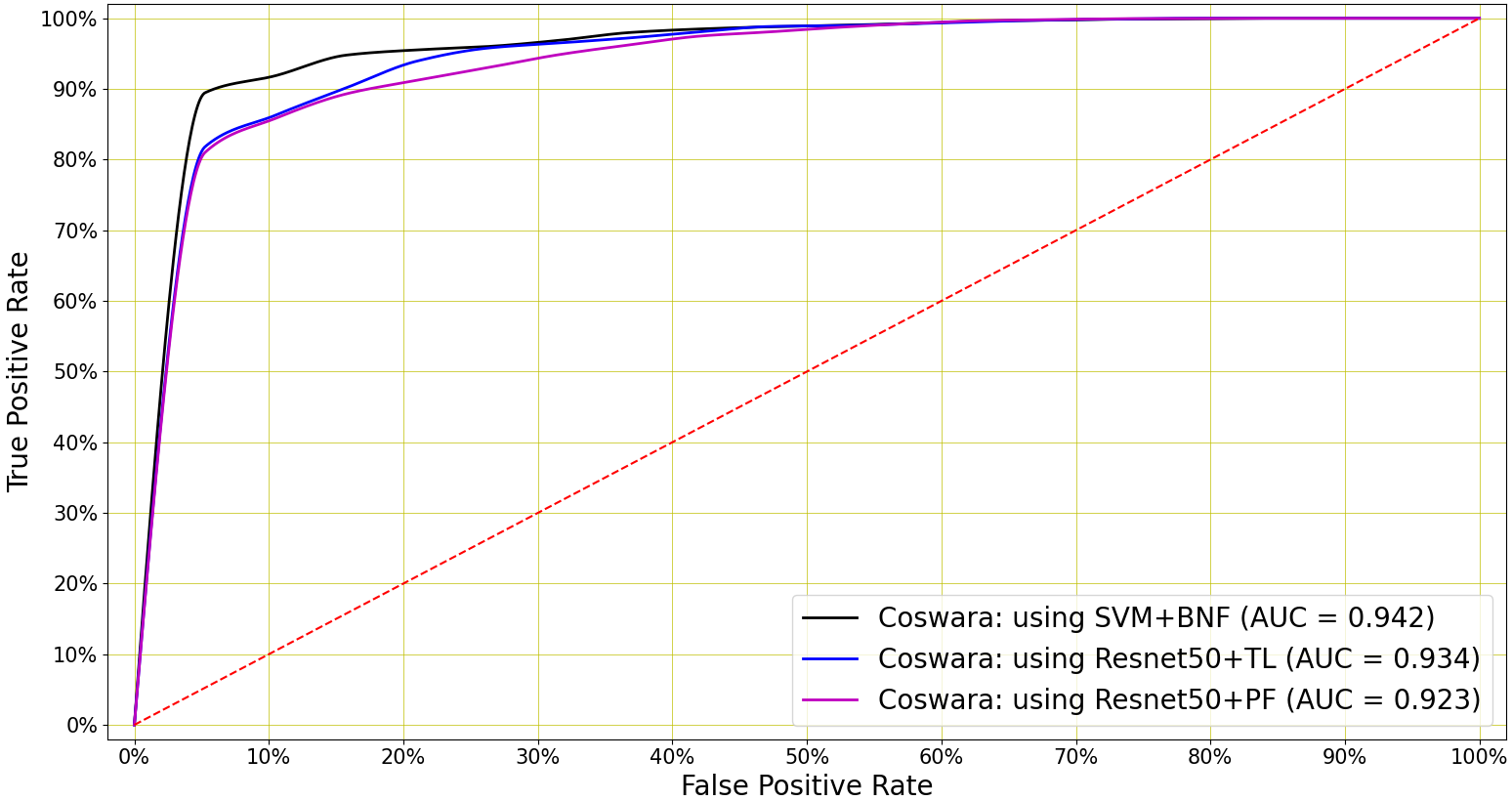}}
	\caption{\textbf{COVID-19 breath classification}: An SVM classifier using bottleneck features (BNF) achieved the highest AUC of 0.942 when classifying COVID-19 breath. The Resnet50 with and without the transfer learning has achieved AUCs of 0.934 and 0.923 respectively, with higher $\sigma_{AUC}$ for the latter (Table \ref{table:breath-classifier-results}). }
	\label{fig:breath-result}
\end{figure}

Table~\ref{table:breath-classifier-results} demonstrates that COVID-19 classification is also possible on the basis of breath signals. 
The baseline systems B8, B9 and B10 are trained directly on the primary features, without pre-training. 
By comparing these baselines with B1, B2 and B3, we see that transfer learning leads to a small improvement in AUC for all three deep architectures. 
Furthermore, systems B4 to B7 show that comparable performance can be achieved by shallow classifiers using the bottleneck features.
The best overall performance (AUC = 0.942) was achieved by an SVM classifier trained on the bottleneck features (system B4).
However, the Resnet50 trained by transfer learning (system B1) performed almost equally well (AUC = 0.934).
The ROC curves for the best-performing COVID-19 breath classifiers are shown in Figure \ref{fig:breath-result}. 
As it was observed for coughs, the standard deviation of the AUC ($\sigma_{AUC}$) is consistently lower for the pre-trained networks.

\begin{figure}[h!]
	\centerline{\includegraphics[width=0.5\textwidth]{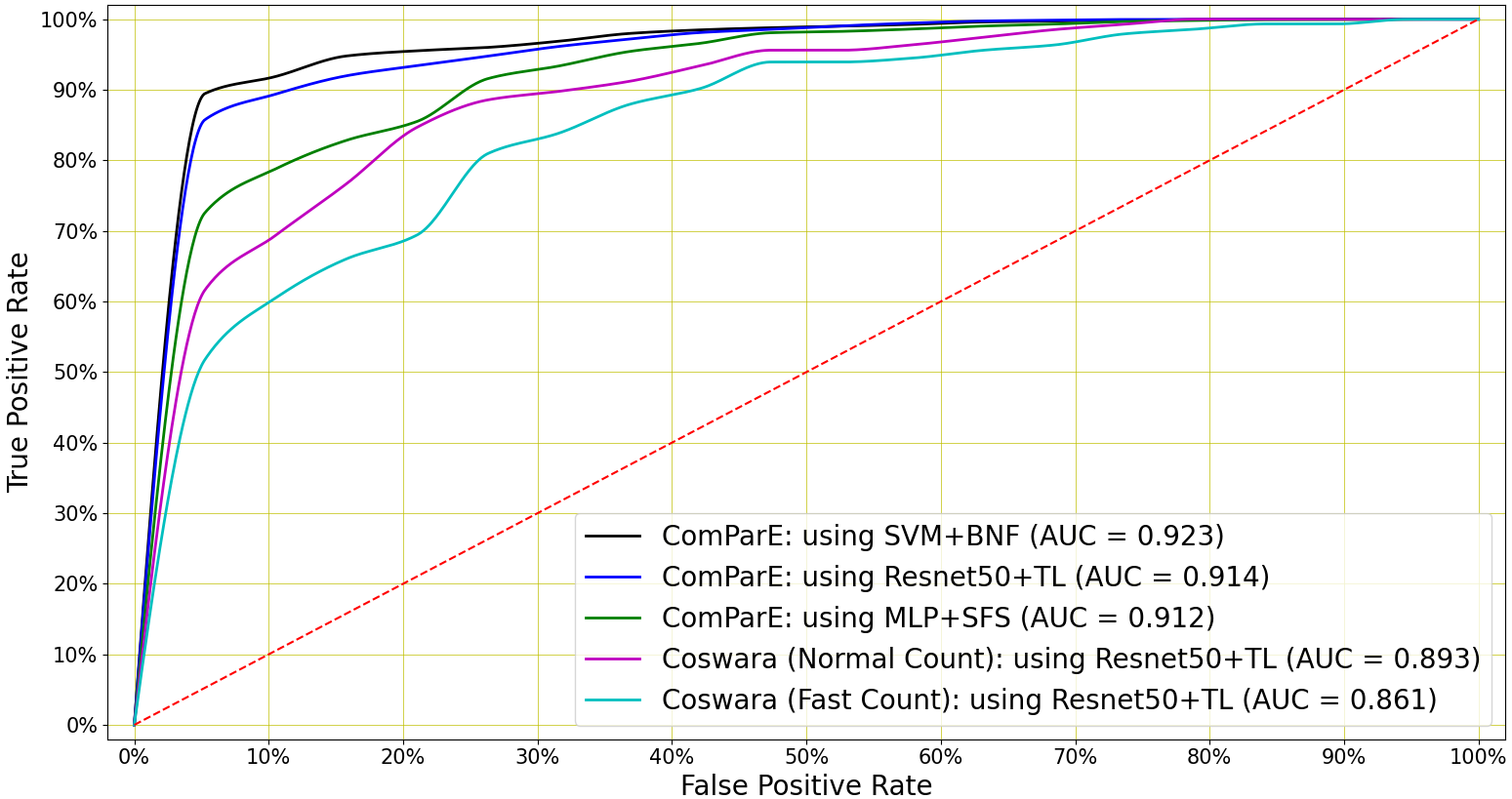}}
	\caption{\textbf{COVID-19 speech classification}: An SVM classifier using bottleneck features (BNF) achieved the highest AUC of 0.923 when classifying COVID-19 speech in ComParE dataset. A Resnet50 trained by transfer learning achieves a slightly lower AUC of 0.914. Speech (normal and fast) in the Coswara dataset can be used to classify COVID-19 with AUCs of 0.893 and 0.861 respectively using a Resnet50 trained by transfer learning. }
	\label{fig:speech-result}
\end{figure}

\subsection{Speech}

\begin{table*}[!h]
	\scriptsize
	\renewcommand\arraystretch{1.4}
	\setlength\minrowclearance{1.0pt}
	\setlength{\tabcolsep}{5pt} 
	\centering
	\caption{\textbf{COVID-19 speech classifier performance:} For the Coswara (fast and normal speech) and the ComParE speech the highest AUCs were 0.893, 0.861 and 0.923 respectively
		and achieved by a Resnet50 trained by transfer learning in the first two cases and an SVM using with bottleneck features in the third case. 
	}
	\begin{tabular}{ c | c | c | c | c | c | c | c | c | c }
		\hline
		\hline
		{\multirow{2}{*}{\textbf{Dataset}}} &
		{\multirow{2}{*}{\textbf{ID}}} &
		{\multirow{2}{*}{\textbf{Classifier}}} & \textbf{Best Feature} &
		\textbf{Best Classifier Hyperparameters} & \multicolumn{5}{c}{\textbf{Performance}} \\
		
		\cline{6-10}
		
		&  &  & \textbf{Hyperparameters} & \textbf{(Optimised inside nested cross-validation)} & \textbf{Spec} & \textbf{Sens} & \textbf{Acc} & \textbf{AUC} & \textbf{$\sigma_{AUC}$} \\
		
		\hline
		\hline
		\multirow{8.5}{*}{Coswara} & \textit{S1} & \textit{Resnet50+TL} & \textit{Table \ref{table:pre-train-hyper-parameter}} & \textit{Default Resnet50 (Table 1 in \cite{he2016deep})} & \textit{90\%} & \textit{85\%} & \textit{87\%} & \textit{0.893} & \textit{3$\times 10^{-3}$} \\
		\cline{2-10}
		\multirow{8}{*}{Normal} & S2 & LSTM+TL & \dittoclosing & Table \ref{table:pre-train-hyper-parameter} & 88\% & 82\% & 85\% & 0.877 & 4$\times 10^{-3}$ \\
		\cline{2-10}
		\multirow{7.5}{*}{Count} & S3 & CNN+TL & \dittoclosing & \dittoclosing & 88\% & 81\% & 85\% & 0.875 & 4$\times 10^{-3}$ \\
		\cline{2-10}
		& S4 & MLP+BNF & \dittoclosing & $\alpha_3$=0.25, $\alpha_7$=60 & 83\% & 85\% & 84\% & 0.871 & 8$\times 10^{-3}$ \\
		\cline{2-10}
		& S5 & SVM+BNF & \dittoclosing & $\alpha_1=10^{-6}$, $\alpha_4=10^{5}$ & 83\% & 85\% & 84\% & 0.867 & 7$\times 10^{-3}$ \\
		\cline{2-10}
		& S6 & KNN+BNF & \dittoclosing & $\alpha_5$=50, $\alpha_6$=10 & 80\% & 85\% & 83\% & 0.868 & 6$\times 10^{-3}$ \\
		\cline{2-10}
		& S7 & LR+BNF & \dittoclosing & $\alpha_1=10^{2}$, $\alpha_2=0.6$, $\alpha_3=0.4$ & 79\% & 83\% & 81\% & 0.852 & 7$\times 10^{-3}$ \\
		\cline{2-10}
		& S8 & Resnet50+PF & $\mathcal{M}=26, \mathcal{F}=2^{10}, \mathcal{S}=120$ & Default Resnet50 (Table 1 in \cite{he2016deep}) & 84\% & 80\% & 82\% & 0.864 & 51$\times 10^{-3}$ \\
		\cline{2-10}
		& S9 & LSTM+PF & $\mathcal{M}=26, \mathcal{F}=2^{11}, \mathcal{S}=150$ & $\beta_3$=0.1, $\beta_4$=32, $\beta_5$=128, $\beta_6$=0.001, $\beta_7$=256, $\beta_8$=170 & 84\% & 78\% & 81\% & 0.844 & 51$\times 10^{-3}$ \\
		\cline{2-10}
		& S10 & CNN+PF & $\mathcal{M}=39, \mathcal{F}=2^{10}, \mathcal{S}=120$ & $\beta_1$=48, $\beta_2$=2, $\beta_3$=0.3, $\beta_4$=32, $\beta_7$=256, $\beta_8$=210 & 82\% & 78\% & 80\% & 0.832 & 52$\times 10^{-3}$ \\
		
		\hline
		
		\multirow{8.5}{*}{Coswara} & \textit{S11} & \textit{Resnet50+TL} & \textit{Table \ref{table:pre-train-hyper-parameter}} & \textit{Default Resnet50 (Table 1 in \cite{he2016deep})} & \textit{84\%} & \textit{78\%} & \textit{81\%} & \textit{0.861} & \textit{2$\times 10^{-3}$} \\
		\cline{2-10}
		\multirow{8}{*}{Fast} & S12 & LSTM+TL & \dittoclosing & Table \ref{table:pre-train-hyper-parameter} & 83\% & 78\% & 81\% & 0.860 & 3$\times 10^{-3}$ \\
		\cline{2-10}
		\multirow{7.5}{*}{Count} & S13 & CNN+TL & \dittoclosing & \dittoclosing & 82\% & 76\% & 79\% & 0.851 & 3$\times 10^{-3}$ \\
		\cline{2-10}
		& S14 & MLP+BNF & \dittoclosing & $\alpha_3$=0.55, $\alpha_7$=70 & 78\% & 83\% & 81\% & 0.858 & 7$\times 10^{-3}$ \\
		\cline{2-10}
		& S15 & SVM+BNF & \dittoclosing & $\alpha_1=10^{4}$, $\alpha_4=10^{-2}$ & 78\% & 83\% & 81\% & 0.856 & 8$\times 10^{-3}$ \\
		\cline{2-10}
		& S16 & KNN+BNF & \dittoclosing & $\alpha_5$=60, $\alpha_6$=15 & 77\% & 83\% & 81\% & 0.854 & 8$\times 10^{-3}$ \\
		\cline{2-10}
		& S17 & LR+BNF & \dittoclosing & $\alpha_1=10^{-3}$, $\alpha_2=0.4$, $\alpha_3=0.6$ & 77\% & 82\% & 80\% & 0.841 & 11$\times 10^{-3}$ \\
		\cline{2-10}
		& S18 & LSTM+PF & $\mathcal{M}=26, \mathcal{F}=2^{11}, \mathcal{S}=120$ & $\beta_3$=0.1, $\beta_4$=32, $\beta_5$=128, $\beta_6$=0.001, $\beta_7$=256, $\beta_8$=170 & 84\% & 80\% & 82\% & 0.856 & 47$\times 10^{-3}$ \\
		\cline{2-10}
		& S19 & Resnet50+PF & $\mathcal{M}=39, \mathcal{F}=2^{10}, \mathcal{S}=150$ & Default Resnet50 (Table 1 in \cite{he2016deep}) & 82\% & 78\% & 80\% & 0.822 & 45$\times 10^{-3}$ \\
		\cline{2-10}
		& S20 & CNN+PF & $\mathcal{M}=52, \mathcal{F}=2^{10}, \mathcal{S}=100$ & $\beta_1$=48, $\beta_2$=2, $\beta_3$=0.3, $\beta_4$=32, $\beta_7$=256, $\beta_8$=210 & 79\% & 77\% & 78\% & 0.810 & 41$\times 10^{-3}$ \\
		
		\hline

		\multirow{12}{*}{ComParE} & S21 & Resnet50+TL & Table \ref{table:pre-train-hyper-parameter} & Default Resnet50 (Table 1 in \cite{he2016deep}) & 84\% & 90\% & 87\% & 0.914 & 4$\times 10^{-3}$ \\
		\cline{2-10}
		& S22 & LSTM+TL & \dittoclosing & Table \ref{table:pre-train-hyper-parameter} & 82\% & 88\% & 85\% & 0.897 & 5$\times 10^{-3}$ \\
		\cline{2-10}
		& S23 & CNN+TL & \dittoclosing & \dittoclosing & 80\% & 88\% & 84\% & 0.892 & 5$\times 10^{-3}$ \\
		\cline{2-10}
		& \textit{S24} & \textit{SVM+BNF} & \dittoclosing & \textit{$\alpha_1=10^{-1}$, $\alpha_4=10^{3}$} & \textit{84\%} & \textit{88\%} & \textit{86\%} & \textit{\textbf{0.923}} & \textit{4$\times 10^{-3}$} \\
		\cline{2-10}
		& S25 & MLP+BNF & \dittoclosing & $\alpha_3$=0.3, $\alpha_7$=60 & 80\% & 88\% & 84\% & 0.905 & 6$\times 10^{-3}$ \\
		\cline{2-10}
		& S26 & KNN+BNF & \dittoclosing & $\alpha_5$=20, $\alpha_6$=15 & 80\% & 86\% & 83\% & 0.891 & 7$\times 10^{-3}$ \\
		\cline{2-10}
		& S27 & LR+BNF & \dittoclosing & $\alpha_1=10^{2}$, $\alpha_2=0.45$, $\alpha_3=0.7$ & 81\% & 85\% & 83\% & 0.890 & 7$\times 10^{-3}$ \\
		\cline{2-10}
		& S28 & MLP+PF+SFS & $\mathcal{M}=26, \mathcal{F}=2^{11}, \mathcal{S}=150$ & $\alpha_3$=0.35, $\alpha_7$=70 & 82\% & 88\% & 85\% & 0.912 & 11$\times 10^{-3}$ \\
		\cline{2-10}
		& S29 & MLP+PF & $\mathcal{M}=26, \mathcal{F}=2^{11}, \mathcal{S}=150$ & $\alpha_3$=0.35, $\alpha_7$=70 & 81\% & 85\% & 83\% & 0.893 & 14$\times 10^{-3}$ \\
		\cline{2-10}
		& S30 & KNN+PF & $\mathcal{B}=100, \mathcal{F}=2^{10}, \mathcal{S}=120$ & $\alpha_5$=70, $\alpha_6$=15 & 80\% & 84\% & 82\% & 0.847 & 16$\times 10^{-3}$ \\
		\cline{2-10}
		& S31 & SVM+PF & $\mathcal{B}=80, \mathcal{F}=2^{11}, \mathcal{S}=120$ & $\alpha_1=10^{-2}$, $\alpha_4=10^{-3}$ & 79\% & 81\% & 80\% & 0.836 & 15$\times 10^{-3}$ \\
		\cline{2-10}
		& S32 & LR+PF & $\mathcal{B}=60, \mathcal{F}=2^{10}, \mathcal{S}=100$ & $\alpha_1=10^{4}$, $\alpha_2=0.35$, $\alpha_3=0.65$ & 69\% & 72\% & 71\% & 0.776 & 18$\times 10^{-3}$ \\

		\hline
		\hline
		
	\end{tabular}
	\label{table:speech-classifier-results}
\end{table*}

Although not as informative as cough or breath audio, COVID-19 classification can also be achieved on the basis of speech audio recordings. 
For Coswara, the best classification performance (AUC = 0.893) was achieved by a Resnet50 after applying transfer learning (system S1). 
For the ComParE data, the top performer (AUC = 0.923) was an SVM trained on the bottleneck features (system S24). 
However, the Resnet50 trained by transfer learning performed almost equally well, with an AUC of 0.914 (system S21). 
Furthermore, while good performance was also achieved when using the deep architectures without applying the transfer learning process (systems S8-S10, S18-S20 and S28-S32), this again was at the cost of a substantially higher standard deviation $\sigma_{AUC}$.
Finally, for the Coswara data,  performance was generally better when speech was uttered at a normal pace rather than a fast pace. 
The ROC curves for the best-performing COVID-19 speech classifiers are shown in 
Figure \ref{fig:speech-result}.

\section{Discussion}\label{sec:discussion}
Previous studies have shown that it is possible to distinguish between the coughing sounds made by COVID-19 positive and COVID-19 negative subjects by means of automatic classification and machine learning. 
However, the fairly small size of datasets with COVID-19 labels limits the effectiveness of these techniques. 
The results of the experiments we have presented in this study show that larger datasets of other vocal and respiratory audio that do not include COVID-19 labels can be leveraged to improve classification performance by applying transfer learning. 
Specifically, we have shown that the accuracy of COVID-19 classification based on coughs can be improved by transfer learning for two datasets (Coswara and Sarcos) while almost optimal performance is achieved on a third dataset (ComParE). 
A similar trend is seen when performing COVID-19 classification based on breath and speech audio. 
However, these two types of audio appear to contain less distinguishing information, since the achieved classification performance is a little lower than it is for cough.
Our best cough classification system has an area under the ROC curve (AUC) of 0.982, despite being trained on what remains a fairly small COVID-19 dataset with 1171 participants (92 COVID-19 positive and 1079 negative). 
Other research reports a similar AUC but using a much larger dataset with 8380 participants (2339 positive and 6041 negative)~\cite{andreu2021generic}. 
Application of transfer learning on all three audio types presented in this study shows better performance than recent studies as well \cite{grant2021rapid}. 
While our experiments also show that shallow classifiers, when used in conjunction with feature selection, can in some cases match or surpass the performance of the deeper architectures; a pre-trained Resnet50 architecture provides consistent optimal or near-optimal performance across all three types of audio signals and datasets. 
Due to the very high computational cost involved, we have not yet applied such feature selection to the deep architectures themselves, and this remains part of our ongoing work.

Another important observation that we can make for all three types of audio signals is 
that transfer learning strongly reduces the variance in the AUC ($\sigma_{AUC}$) exhibited by the deep classifiers during cross-validation (Table \ref{table:cough-classifier-results}, \ref{table:breath-classifier-results} and \ref{table:speech-classifier-results}). 
This suggests that transfer learning leads to more consistent classifiers that are less prone to over-fitting and better able to generalise on the unseen data. 
This is important because robustness to variable testing conditions is essential in implementing COVID-19 classification as a method of screening.

An informal listening assessment of the Coswara and the ComParE data indicates that the former has greater variance and more noise than the latter. 
Our experimental results presented in Table \ref{table:cough-classifier-results}, \ref{table:breath-classifier-results} and \ref{table:speech-classifier-results} found that, for speech classification on a noisy data, fine-tuning the pre-trained networks demonstrates better performance, while for cleaner data, extracting bottleneck features and then applying a shallow classifier exhibits better performance.  
It is interesting to note that MFCCs are always the features of choice for this noisier dataset, while the log energies of linear filters are often preferred for the less noisy data. 
Although all other classifiers have shown the best performance when using these log-filterbank energy features, MLP has always performed the best on MFCCs and has been proved to be the best classifier in classifying COVID-19 speech.
A similar conclusion was also drawn in~\cite{botha2018detection}, where coughs were recorded in a controlled environment with little environmental noise.
A higher number of segments also generally leads to better performance as it allows the classifier to find more detailed temporal patterns in the audio signal.


\section{Conclusions}\label{sec:conclusion}

In this study, we have demonstrated that transfer learning can be used to improve the performance and robustness of the DNN classifiers for COVID-19 detection in vocal audio such as cough, breath and speech. 
We have used a 10.29 hour audio data corpus, which do not have any COVID-19 labels, to pre-train a CNN, an LSTM and a Resnet50. 
This data contains four classes: cough, sneeze, speech and noise. 
In addition, we have used the same architectures to extract bottleneck features by removing the final layers from the pre-trained models. 
Three smaller datasets containing cough, breath and speech audio with COVID-19 labels were then used to fine-tune the pre-trained COVID-19 audio classifiers using nested leave-$p$-out cross-validation. 
Our results show that a pre-trained Resnet50 classifier that is either fine-tuned or used as a bottleneck extractor delivers optimal or near-optimal performance across all datasets and all three audio classes. 
The results show that transfer learning using the larger dataset without COVID-19 labels led not only to improved performance, but also to a smaller standard deviation of the classifier AUC, indicating better generalisation. 
The use of bottleneck features, which are extracted by the pre-trained deep models and therefore also a way of incorporating out-of-domain data, also provided a reduction in this standard deviation and near-optimal performance. 
The experiments show that cough audio carries the strongest COVID-19 signatures, followed by breath and speech.
The best-performing COVID-19 classifier achieved an area under the ROC curve (AUC) of 0.982 for cough, followed by an AUC of 0.942 for breath and 0.923 for speech. 

Finally, we note that, for the shallow classifiers, hyperparameter optimisation selected a higher number of MFCCs and also a more densely populated filterbank than what is required to match the resolution of the human auditory system.
This agrees with an observation we have already made in our previous work that the information used by the classifiers to detect COVID-19 signature is at least to some extent not perceivable by the human ear. 

We conclude that successful classification is possible for all three classes of audio considered. 
However, deep transfer learning improves COVID-19 detection on the basis of cough, breath and speech signals, yielding automatic classifiers with higher accuracies and greater robustness.
This is significant since such COVID-19 screening is inexpensive, easily deployable, non-contact and does not require medical expertise or laboratory facilities.
Therefore it has the potential to decrease the load on the health care systems.

As a part of ongoing work, we are considering the application of feature selection in the deep architectures, the fusion of classifiers using various audio classes like cough, breath and speech, as well as the optimisation and adaptation necessary to allow deployment on a smartphone or similar mobile platform.


\section{Acknowledgements}
This research was supported by the South African Medical Research Council (SAMRC) through its Division of Research Capacity Development under the Research Capacity Development Initiative as well as the COVID-19 IMU EMC allocation from funding received from the South African National Treasury. 
Support was also received from the European Union through the EDCTP2 programme (TMA2017CDF-1885). The content and findings reported are the sole deduction, view and responsibility of the researcher and do not reflect the official position and sentiments of the SAMRC or the EDCTP. 

We would also like to thank South African Centre for High Performance Computing (CHPC) for providing computational resources on their Lengau cluster to support this research, and gratefully acknowledge the support of Telkom South Africa.

We also especially thank Igor Miranda, Corwynne Leng, Renier Botha, Jordan Govendar and Rafeeq du Toit for their support in data collection and annotation.


\bibliographystyle{IEEEbib}
\bibliography{reference}

\begin{thebibliography}{10}

\bibitem{carfi2020persistent}
Angelo Carf{\`\i}, Roberto Bernabei, Francesco Landi, et~al.,
\newblock ``Persistent symptoms in patients after acute {COVID}-19,''
\newblock {\em {JAMA}}, vol. 324, no. 6, pp. 603--605, 2020.

\bibitem{wang2020clinical}
Dawei Wang, Bo~Hu, Chang Hu, Fangfang Zhu, Xing Liu, Jing Zhang, Binbin Wang,
  Hui Xiang, Zhenshun Cheng, Yong Xiong, et~al.,
\newblock ``Clinical characteristics of 138 hospitalized patients with 2019
  novel coronavirus--infected pneumonia in {W}uhan, {C}hina,''
\newblock {\em {JAMA}}, vol. 323, no. 11, pp. 1061--1069, 2020.

\bibitem{marini2020management}
John~J Marini and Luciano Gattinoni,
\newblock ``Management of {COVID}-19 respiratory distress,''
\newblock {\em JAMA}, vol. 323, no. 22, pp. 2329--2330, 2020.

\bibitem{aguiar2020inside}
Diego Aguiar, Johannes~Alexander Lobrinus, Manuel Schibler, Tony Fracasso, and
  Christelle Lardi,
\newblock ``Inside the lungs of {COVID}-19 disease,''
\newblock {\em International Journal of Legal Medicine}, vol. 134, pp.
  1271--1274, 2020.

\bibitem{ziehr2020respiratory}
David~R Ziehr, Jehan Alladina, Camille~R Petri, Jason~H Maley, Ari Moskowitz,
  Benjamin~D Medoff, Kathryn~A Hibbert, B~Taylor Thompson, and C~Corey Hardin,
\newblock ``Respiratory pathophysiology of mechanically ventilated patients
  with {COVID}-19: a cohort study,''
\newblock {\em American Journal of Respiratory and Critical Care Medicine},
  vol. 201, no. 12, pp. 1560--1564, 2020.

\bibitem{davis2021breath}
Cristina~E Davis, Michael Schivo, and Nicholas~J Kenyon,
\newblock ``A breath of fresh air--the potential for {COVID}-19 breath
  diagnostics,''
\newblock {\em EBioMedicine}, vol. 63, 2021.

\bibitem{grassin2021metabolomics}
Stanislas Grassin-Delyle, Camille Roquencourt, Pierre Moine, Gabriel Saffroy,
  Stanislas Carn, Nicholas Heming, J{\'e}r{\^o}me Fleuriet, H{\'e}l{\`e}ne
  Salvator, Emmanuel Naline, Louis-Jean Couderc, et~al.,
\newblock ``Metabolomics of exhaled breath in critically ill {COVID}-19
  patients: A pilot study,''
\newblock {\em EBioMedicine}, vol. 63, pp. 103154, 2021.

\bibitem{ruszkiewicz2020diagnosis}
Dorota~M Ruszkiewicz, Daniel Sanders, Rachel O'Brien, Frederik Hempel,
  Matthew~J Reed, Ansgar~C Riepe, Kenneth Bailie, Emma Brodrick, Kareen
  Darnley, Richard Ellerkmann, et~al.,
\newblock ``Diagnosis of {COVID}-19 by analysis of breath with gas
  chromatography-ion mobility spectrometry-a feasibility study,''
\newblock {\em EClinicalMedicine}, vol. 29, pp. 100609, 2020.

\bibitem{walvekar2020detection}
Sanika Walvekar, Dr~Shinde, et~al.,
\newblock ``Detection of {COVID}-19 from {CT} images using {R}esnet50,''
\newblock in {\em 2nd International Conference on Communication \& Information
  Processing ({ICCIP}) 2020﻿}, May 2020.

\bibitem{sotoudeh2020artificial}
Houman Sotoudeh, Mohsen Tabatabaei, Baharak Tasorian, Kamran Tavakol, Ehsan
  Sotoudeh, and Abdol~Latif Moini,
\newblock ``Artificial {I}ntelligence {E}mpowers {R}adiologists to
  {D}ifferentiate {P}neumonia {I}nduced by {COVID}-19 versus {I}nfluenza
  {V}iruses,''
\newblock {\em Acta Informatica Medica}, vol. 28, no. 3, pp. 190, 2020.

\bibitem{yildirim2020deep}
Muhammed Yildirim and Ahmet Cinar,
\newblock ``A {D}eep {L}earning {B}ased {H}ybrid {A}pproach for {COVID}-19
  {D}isease {D}etections,''
\newblock {\em Traitement du Signal}, vol. 37, no. 3, pp. 461--468, 2020.

\bibitem{higenbottam2002chronic}
T~Higenbottam,
\newblock ``Chronic cough and the cough reflex in common lung diseases,''
\newblock {\em Pulmonary {P}harmacology \& {T}herapeutics}, vol. 15, no. 3, pp.
  241--247, 2002.

\bibitem{chang2008chronic}
AB~Chang, GJ~Redding, and ML~Everard,
\newblock ``Chronic wet cough: protracted bronchitis, chronic suppurative lung
  disease and bronchiectasis,''
\newblock {\em Pediatric {P}ulmonology}, vol. 43, no. 6, pp. 519--531, 2008.

\bibitem{chung2008prevalence}
Kian~Fan Chung and Ian~D Pavord,
\newblock ``Prevalence, pathogenesis, and causes of chronic cough,''
\newblock {\em The Lancet}, vol. 371, no. 9621, pp. 1364--1374, 2008.

\bibitem{knocikova2008wavelet}
J~Knocikova, J~Korpas, M~Vrabec, and M~Javorka,
\newblock ``Wavelet analysis of voluntary cough sound in patients with
  respiratory diseases,''
\newblock {\em Journal of Physiology and Pharmacology}, vol. 59, no. Suppl 6,
  pp. 331--40, 2008.

\bibitem{imran2020ai4covid}
Ali Imran, Iryna Posokhova, Haneya~N Qureshi, Usama Masood, Muhammad~Sajid
  Riaz, Kamran Ali, Charles~N John, MD~Iftikhar Hussain, and Muhammad Nabeel,
\newblock ``{AI4COVID}-19: {AI} enabled preliminary diagnosis for {COVID}-19
  from cough samples via an app,''
\newblock {\em Informatics in Medicine Unlocked}, vol. 20, pp. 100378, 2020.

\bibitem{laguarta2020covid}
Jordi Laguarta, Ferran Hueto, and Brian Subirana,
\newblock ``{COVID}-19 {A}rtificial {I}ntelligence {D}iagnosis using only
  {C}ough {R}ecordings,''
\newblock {\em IEEE Open Journal of Engineering in Medicine and Biology}, vol.
  1, pp. 275--281, 2020.

\bibitem{brown2020exploring}
Chlo{\"e} Brown, Jagmohan Chauhan, Andreas Grammenos, Jing Han, Apinan
  Hasthanasombat, Dimitris Spathis, Tong Xia, Pietro Cicuta, and Cecilia
  Mascolo,
\newblock ``Exploring {A}utomatic {D}iagnosis of {COVID}-19 from {C}rowdsourced
  {R}espiratory {S}ound {D}ata,''
\newblock in {\em Proceedings of the 26th ACM SIGKDD International Conference
  on Knowledge Discovery \& Data Mining}, 2020, pp. 3474--3484.

\bibitem{coppock2021end}
Harry Coppock, Alex Gaskell, Panagiotis Tzirakis, Alice Baird, Lyn Jones, and
  Bj{\"o}rn Schuller,
\newblock ``End-to-end convolutional neural network enables {COVID}-19
  detection from breath and cough audio: a pilot study,''
\newblock {\em BMJ Innovations}, vol. 7, no. 2, 2021.

\bibitem{pahar2020covid}
Madhurananda Pahar, Marisa Klopper, Robin Warren, and Thomas Niesler,
\newblock ``{COVID}-19 cough classification using machine learning and global
  smartphone recordings,''
\newblock {\em Computers in Biology and Medicine}, vol. 135, pp. 104572, 2021.

\bibitem{sharma2020coswara}
Neeraj Sharma, Prashant Krishnan, Rohit Kumar, Shreyas Ramoji, Srikanth~Raj
  Chetupalli, Nirmala R., Prasanta~Kumar Ghosh, and Sriram Ganapathy,
\newblock ``Coswara--{A} {D}atabase of {B}reathing, {C}ough, and {V}oice
  {S}ounds for {COVID}-19 {D}iagnosis,''
\newblock in {\em Proc. Interspeech 2020}, 2020, pp. 4811--4815.

\bibitem{Schuller21-TI2}
Bj{\"o}rn~W.\ Schuller, Anton Batliner, Christian Bergler, Cecilia Mascolo,
  Jing Han, Iulia Lefter, Heysem Kaya, Shahin Amiriparian, Alice Baird, Lukas
  Stappen, Sandra Ottl, Maurice Gerczuk, Panaguiotis Tzirakis, Chloë Brown,
  Jagmohan Chauhan, Andreas Grammenos, Apinan Hasthanasombat, Dimitris Spathis,
  Tong Xia, Pietro Cicuta, M.\~Rothkrantz Leon~J.\, Joeri Zwerts, Jelle Treep,
  and Casper Kaandorp,
\newblock ``{The {INTERSPEECH} 2021 {C}omputational {P}aralinguistics
  {C}hallenge: {COVID}-19 {C}ough, {COVID}-19 {S}peech, {E}scalation \&
  {P}rimates},''
\newblock in {\em {Proceedings INTERSPEECH 2021, 22nd Annual Conference of the
  International Speech Communication Association}}, Brno, Czechia, September
  2021, ISCA,
\newblock to appear.

\bibitem{pahar2021deep}
Madhurananda Pahar, Igor Miranda, Andreas Diacon, and Thomas Niesler,
\newblock ``Deep {N}eural {N}etwork based {C}ough {D}etection using
  {B}ed-mounted {A}ccelerometer {M}easurements,''
\newblock in {\em ICASSP 2021 - 2021 IEEE International Conference on
  Acoustics, Speech and Signal Processing (ICASSP)}, 2021, pp. 8002--8006.

\bibitem{botha2018detection}
GHR Botha, G~Theron, RM~Warren, M~Klopper, K~Dheda, PD~Van~Helden, and
  TR~Niesler,
\newblock ``Detection of {T}uberculosis by {A}utomatic {C}ough {S}ound
  {A}nalysis,''
\newblock {\em Physiological Measurement}, vol. 39, no. 4, pp. 045005, 2018.

\bibitem{pahar2021tb}
Madhurananda Pahar, Marisa Klopper, Byron Reeve, Robin Warren, Grant Theron,
  and Thomas Niesler,
\newblock ``Automatic {C}ough {C}lassification for {T}uberculosis {S}creening
  in a {R}eal-{W}orld {E}nvironment,''
\newblock {\em arXiv preprint arXiv:2103.13300}, 2021.

\bibitem{gemmeke2017audio}
Jort~F Gemmeke, Daniel~PW Ellis, Dylan Freedman, Aren Jansen, Wade Lawrence,
  R~Channing Moore, Manoj Plakal, and Marvin Ritter,
\newblock ``Audio set: {A}n ontology and human-labeled dataset for audio
  events,''
\newblock in {\em 2017 IEEE International Conference on Acoustics, Speech and
  Signal Processing (ICASSP)}. IEEE, 2017, pp. 776--780.

\bibitem{font2013freesound}
Frederic Font, Gerard Roma, and Xavier Serra,
\newblock ``Freesound technical demo,''
\newblock in {\em Proceedings of the 21st ACM International Conference on
  Multimedia}, 2013, pp. 411--412.

\bibitem{miranda2019comparative}
Igor~DS Miranda, Andreas~H Diacon, and Thomas~R Niesler,
\newblock ``A comparative study of features for acoustic cough detection using
  deep architectures,''
\newblock in {\em 2019 41st Annual International Conference of the IEEE
  Engineering in Medicine and Biology Society (EMBC)}. IEEE, 2019, pp.
  2601--2605.

\bibitem{panayotov2015librispeech}
Vassil Panayotov, Guoguo Chen, Daniel Povey, and Sanjeev Khudanpur,
\newblock ``Librispeech: {A}n {ASR} corpus based on public domain audio
  books,''
\newblock in {\em 2015 IEEE international conference on acoustics, speech and
  signal processing (ICASSP)}. IEEE, 2015, pp. 5206--5210.

\bibitem{van2007experimental}
Jason Van~Hulse, Taghi~M Khoshgoftaar, and Amri Napolitano,
\newblock ``Experimental perspectives on learning from imbalanced data,''
\newblock in {\em Proceedings of the 24th International Conference on Machine
  Learning}, 2007, pp. 935--942.

\bibitem{krawczyk2016learning}
Bartosz Krawczyk,
\newblock ``Learning from imbalanced data: open challenges and future
  directions,''
\newblock {\em Progress in Artificial Intelligence}, vol. 5, no. 4, pp.
  221--232, 2016.

\bibitem{chawla2002smote}
Nitesh~V Chawla, Kevin~W Bowyer, Lawrence~O Hall, and W~Philip Kegelmeyer,
\newblock ``{SMOTE}: synthetic minority over-sampling technique,''
\newblock {\em Journal of {A}rtificial {I}ntelligence {R}esearch}, vol. 16, pp.
  321--357, 2002.

\bibitem{muguli2021dicova}
Ananya Muguli, Lancelot Pinto, Neeraj Sharma, Prashant Krishnan, Prasanta~Kumar
  Ghosh, Rohit Kumar, Shreyas Ramoji, Shrirama Bhat, Srikanth~Raj Chetupalli,
  Sriram Ganapathy, et~al.,
\newblock ``Di{COVA} {C}hallenge: {D}ataset, task, and baseline system for
  {COVID}-19 diagnosis using acoustics,''
\newblock {\em arXiv preprint arXiv:2103.09148}, 2021.

\bibitem{bachu2010voiced}
RG~Bachu, S~Kopparthi, B~Adapa, and Buket~D Barkana,
\newblock ``Voiced/unvoiced decision for speech signals based on zero-crossing
  rate and energy,''
\newblock in {\em Advanced {T}echniques in {C}omputing {S}ciences and
  {S}oftware {E}ngineering}, pp. 279--282. Springer, 2010.

\bibitem{pahar_coding_2020}
Madhurananda Pahar and Leslie~S Smith,
\newblock ``Coding and {D}ecoding {S}peech using a {B}iologically {I}nspired
  {C}oding {S}ystem,''
\newblock in {\em 2020 IEEE Symposium Series on Computational Intelligence
  (SSCI)}. IEEE, 2020, pp. 3025--3032.

\bibitem{chatrzarrin2011feature}
Hanieh Chatrzarrin, Amaya Arcelus, Rafik Goubran, and Frank Knoefel,
\newblock ``Feature extraction for the differentiation of dry and wet cough
  sounds,''
\newblock in {\em IEEE International Symposium on Medical Measurements and
  Applications}. IEEE, 2011, pp. 162--166.

\bibitem{alsabek2020studying}
Mohammed~Bader Alsabek, Ismail Shahin, and Abdelfatah Hassan,
\newblock ``Studying the {S}imilarity of {COVID}-19 {S}ounds based on
  {C}orrelation {A}nalysis of {MFCC},''
\newblock in {\em 2020 International Conference on Communications, Computing,
  Cybersecurity, and Informatics (CCCI)}. IEEE, 2020, pp. 1--5.

\bibitem{aydin2009log}
Serap Ayd{\i}n, Hamdi~Melih Sarao{\u{g}}lu, and Sad{\i}k Kara,
\newblock ``Log energy entropy-based {EEG} classification with multilayer
  neural networks in seizure,''
\newblock {\em Annals of {B}iomedical {E}ngineering}, vol. 37, no. 12, pp.
  2626, 2009.

\bibitem{he2016deep}
Kaiming He, Xiangyu Zhang, Shaoqing Ren, and Jian Sun,
\newblock ``Deep residual learning for image recognition,''
\newblock in {\em Proceedings of the IEEE {C}onference on {C}omputer {V}ision
  and {P}attern {R}ecognition}, 2016, pp. 770--778.

\bibitem{silnova2018but}
Anna Silnova, Pavel Matejka, Ondrej Glembek, Oldrich Plchot, Ondrej
  Novotn{\`y}, Frantisek Grezl, Petr Schwarz, Lukas Burget, and Jan
  Cernock{\`y},
\newblock ``{BUT}/{P}honexia {B}ottleneck {F}eature {E}xtractor.,''
\newblock in {\em Odyssey}, 2018, pp. 283--287.

\bibitem{song2015deep}
Yan Song, Ian McLoughLin, and Lirong Dai,
\newblock ``Deep {B}ottleneck {F}eature for {I}mage {C}lassification,''
\newblock in {\em Proceedings of the 5th ACM on International Conference on
  Multimedia Retrieval}, 2015, pp. 491--494.

\bibitem{nguyen2013optimizing}
Quoc~Bao Nguyen, Jonas Gehring, Kevin Kilgour, and Alex Waibel,
\newblock ``Optimizing deep bottleneck feature extraction,''
\newblock in {\em The 2013 RIVF International Conference on Computing \&
  Communication Technologies-Research, Innovation, and Vision for Future
  (RIVF)}. IEEE, 2013, pp. 152--156.

\bibitem{liu2019leave}
Shiqin Liu,
\newblock ``Leave-$ p $-{O}ut {C}ross-{V}alidation {T}est for {U}ncertain
  {V}erhulst-{P}earl {M}odel {W}ith {I}mprecise {O}bservations,''
\newblock {\em IEEE Access}, vol. 7, pp. 131705--131709, 2019.

\bibitem{fawcett2006introduction}
Tom Fawcett,
\newblock ``An introduction to {ROC} analysis,''
\newblock {\em Pattern Recognition Letters}, vol. 27, no. 8, pp. 861--874,
  2006.

\bibitem{devijver1982pattern}
Pierre~A Devijver and Josef Kittler,
\newblock {\em Pattern recognition: A statistical approach},
\newblock Prentice Hall, 1982.

\bibitem{andreu2021generic}
Javier Andreu-Perez, Humberto P{\'e}rez-Espinosa, Eva Timonet, Mehrin Kiani,
  Manuel~Ivan Giron-Perez, Alma~B Benitez-Trinidad, Delaram Jarchi, Alejandro
  Rosales, Nick Gkatzoulis, Orion~F Reyes-Galaviz, et~al.,
\newblock ``A {G}eneric {D}eep {L}earning {B}ased {C}ough {A}nalysis {S}ystem
  from {C}linically {V}alidated {S}amples for {P}oint-of-{N}eed {C}ovid-19
  {T}est and {S}everity {L}evels,''
\newblock {\em IEEE Transactions on Services Computing}, pp. 1--1, 2021.

\bibitem{grant2021rapid}
Drew Grant, Ian McLane, and James West,
\newblock ``Rapid and {S}calable {COVID}-19 {S}creening using {S}peech,
  {B}reath, and {C}ough {R}ecordings,''
\newblock in {\em 2021 IEEE EMBS International Conference on Biomedical and
  Health Informatics (BHI)}. IEEE, 2021, pp. 1--6.

\end{thebibliography}

%
%

\end{document}